\documentclass[superscriptaddress,prd,amsfonts,aps,nofootinbib,notitlepage,10pt]{revtex4-1}
\pdfoutput=1
\usepackage{amsmath,amssymb}
\usepackage{graphicx}
\usepackage[utf8]{inputenc}
\usepackage{cancel}
\usepackage[colorlinks=true]{hyperref}
\usepackage{color}
\usepackage{soul}
\usepackage{multirow}
\usepackage{slashed}
\usepackage{enumitem}
\newcommand{\nn}{\nonumber}

\begin{document}
\title{Multicomponent scalar dark matter at high-intensity proton beam experiments}
\author{Amalia Betancur}
\email{amalia.betancur@eia.edu.co}
\affiliation{
\textit{\small  Grupo F\'{i}sica Te\'{o}rica y Aplicada, Universidad EIA,} \\
\textit{\small  A.A. 7516, Envigado, Colombia}\\[4mm]
}

\author{Andr\'es Castillo}
  \email{acastillo@iac.es}
  \affiliation{\textit{\small Grupo de F\'{i}sica Experimental, Big Data y Anal\'itica, Universidad Sergio Arboleda,}\\
 \textit{\small  Bogot\'a DC, Colombia}\\[4mm]
 }
 \affiliation{\textit{\small 
Instituto de Astrof\'isica de Canarias, C/ V\'ia L\'actea, s/n E38205}\\
 \textit{\small  La Laguna (Tenerife), Spain}\\[4mm]
 }
 \affiliation{\textit{\small 
Universidad de La Laguna, Departamento de Astrof\'isica, La Laguna, Tenerife, Spain}\\
 \textit{\small  La Laguna (Tenerife), Spain}\\[4mm]
 }
  \author{Guillermo Palacio}
    \email{guillermo.palacio38@eia.edu.co}
\affiliation{
\textit{\small  Grupo F\'{i}sica Te\'{o}rica y Aplicada, Universidad EIA,} \\
\textit{\small  A.A. 7516, Envigado, Colombia}\\[4mm]
}
\author{Juan Suarez}
    \email{juan.suarez47@eia.edu.co}
\affiliation{
\textit{\small  Grupo F\'{i}sica Te\'{o}rica y Aplicada, Universidad EIA,} \\
\textit{\small  A.A. 7516, Envigado, Colombia}\\[4mm]
}

\begin{abstract}

  We study a scalar dark matter (DM) model with two DM species coupled to the Standard Model (SM) particles via a sub-GeV dark photon. In this model, we find that DM conversion occurs through the dark photon and it plays a fundamental role in setting the observed relic abundance. Furthermore, the two DM candidates can be produced at fixed-target experiments a la Beam-Dump. Detailed predictions for signal and backgrounds are obtained with the help of MadDump and NuWro Montecarlo generators. We explore the potential reach on the sensitivity of DUNE near detector and SHiP experiment, and we find that portions of the parameter space will be within reach of the two experiments.
\end{abstract}

\maketitle
\section{Introduction}
\label{sec:introduction}

Despite the increasing evidence of the existence of dark matter (DM), its direct observation continues to be elusive. Many searches for DM have been carried out in a variety of experiments, in indirect detection looking for signals of DM annihilation in outer space, underground facilities looking for DM interaction with a detector, and even at the Large Hadron Collider, all yielding null results so far. The DM program has mostly focused on the search for a DM particle at the electroweak scale, because it has been considered that such mass scale is natural for a Weakly Interactive Massive Particle (WIMP), one of the preferred DM candidates. Nevertheless, the null results so far, have prompted the community to consider different candidates and/or mass ranges. For instance, DM candidate could still be a WIMP with thermal decoupling but with a mass that lies well below the electroweak scale and is therefore not within reach of the usual direct detection experiments sensitive to weak scale WIMPS. It is then natural that such DM candidate has escaped observation. 

For this DM to communicate with the SM a new portal is needed, which could be scalar, vector or fermion \cite{Batell:2009di}. One of the most widely studied models is the vector portal, where the mediator is a new gauge boson, similar to the photon, which communicates to the SM through
a kinetic mixing \cite{Pospelov:2007mp,Pospelov:2008zw,Fabbrichesi:2020wbt,Lee:2015gsa,
  Feldman:2007wj,Hook:2010tw,Frandsen:2011cg,Arcadi:2018tly,CortinaGil:2019nuo,
  Naaz:2019pvs,Lao:2020inc,Chun:2010ve,Babu:1997st}.
Such new gauge boson, the dark photon, has previously been proposed to explain the observation of gamma-ray lines in the center of the Milky Way \cite{Pospelov:2007mp} and the discrepancy between the theoretical prediction and the measured value of the muon anomalous magnetic moment $\Delta a_{\mu}$ \cite{Abi:2021gix,Mohlabeng:2019vrz}, as is included in the historical account of \cite{Filippi:2020kii}. 
In this work we will consider a dark photon that arises from a $U(1)_D$ symmetry, which is broken, hence yielding a massive vector boson. This dark photon will not be considered for the aforementioned reported anomalies, but as a portal to communicate the DM sector with the SM sector.

In regards to the DM particles that communicate with the SM through the dark photon, the preferred candidates have been singlets, either fermion or scalar particles that are only charged under the $U(1)_D$ symmetry. Nevertheless, it is important to consider models that go beyond the simplest scenarios and that have more potential for rich phenomenology. For instance, the total relic abundance could be saturated by several DM candidates,  which is natural considering that 5$\%$ of the matter-energy content of the Universe is made of a myriad of particles. This idea, dubbed multicomponent DM has gained more attention because it introduces viable candidates while also allowing for a rich phenomenology that could be within reach of actual and near-future experiments. Such models allow for different parameter space than the one-component and may include important effects such as DM conversion. Moreover, we know DM is multicomponent because active neutrinos make up a small percentage of the DM \cite{Primack:2001ib}. Thus, it makes sense to consider more than one DM particle.

Facilities that rely on high-intensity proton collisions have a great potential to look for dark-sector models, because in the energetic collisions it is possible to produce the dark sector particles which could be detected through interactions such as electron recoil. Those interactions have the potential to probe low masses, and although there could be a large background coming from neutrinos, it is also possible to disentangle the two types of events. This is the case of DUNE (Deep Underground Neutrino Experiment)~\cite{Abi:2020wmh} and SHiP (Search for Hidden Particles)~\cite{Shirobokov:2021bno} experiments.

For instance, DUNE-LBNF (Long-Baseline Neutrino Facility) will use a high-intensity proton beam of 120 GeV producing a flux of neutrinos with energies in the range of several hundreds of MeVs up to several GeVs. DUNE will also have two state-of-the-art near and far detectors, composed mainly of liquid argon. For the near detector (ND), the Precision Reaction of Independent Spectrum Measurement (PRISM) concept incorporates a moving complex across different off-axis positions~\cite{CDRND2021}.
The PRISM idea allows to control systematically the uncertainties from neutrino flux measurements and neutrino-nucleus interactions (i.e. cross sections constraints); and thus improve the precision tests for neutrino oscillations~\cite{Marshall_2020}. Besides, as we will show in this paper, the PRISM concept leads to a background reduction, converting the DUNE ND into a crucial experiment to search for dark-sector particles produced after the collision of the proton beam with the target.

In this work, we consider a DM model where two singlet scalars are the DM candidates, both charged under the $U(1)_D$ symmetry and under discrete symmetries that ensure the stability of both DM candidates. The relic density phenomenology is presented, including the impact of DM conversion. 
In addition, we  investigate
the potential reach on the sensitivity
of the two-component scalar dark matter model.
For that aim, we consider two currently under construction experiments:
 DUNE and SHiP. In particular, we
consider a dark photon in the mass window of $10-900$ MeV
with on-shell decay to scalar DM and
study the expected sensitivity of the experiment on the
parameter space of the model.
For the construction of the sensitivity,
we compare the neutrino-electron ($\nu-e$) scattering
(the background) against the DM-electron (DM$-e$) scattering
(the signal). 
  Our results are novel and go beyond the analyses presented in
  Refs.~\cite{DeRomeri:2019kic,Celentano:2020vtu,
  Breitbach:2021gvv,Buonocore:2019esg} in the following ways:

\begin{itemize}
  
\item We propose a model with two dark matter components,
  leading to new annihilation channels, impacting the relic density.
  Additionally, we perform a comprehensive exploration of the parameter space
  of the model given by the relic abundance observable.
  
\item Regarding the background studies, we show that the {\tt NuWro}
  Monte Carlo generator is suitable to simulate neutrino-induced events
  in the DUNE-ND. 
 
\item Moreover, we show that the SHiP experiment
  is crucial in obtaining
  the most stringent constraints 
  of the sensitivity  for the multicomponent model.
  
\item  For two dark matter component, DUNE-ND and
  SHiP experiments allow us to explore
  different parameter space  compared to the one
  allowed for the one dark matter candidate model.

\end{itemize}

This paper is organized as follows: In Sec.~\ref{sec:model}
we introduce the  model Lagrangian, particle
content and list the tools used
in the phenomenological study for this work.
In Sec.~\ref{DMpheno} we present
the constraints on DM observables,
discuss DM conversion, and investigate the
 phenomenological implications
of  two dark matter candidates.
Sec.~\ref{sec:dm@nu-facilities}
discusses the experimental set-up
for light
DM production at high-intensity proton beam experiments,
and displays some relevant results
regarding the background and signal
simulation. Additionally, we show 
the advantages of moving the DUNE ND
to off-axis configurations.
We also present the potential
reach on the sensitivity for
DUNE and SHiP in Sec.~\ref{sec:sensitivity_analysis}.
 Finally, our results are summarized in Sec.~\ref{sec:conclusions}.


\section{The Model}
\label{sec:model}
We consider an extension of the SM by adding two new gauge singlets to the scalar sector. One complex singlet is charged under a new global symmetry $Z_2$ and the other one is charged under a $Z_2'$, while the SM particles are uncharged. These symmetries will remain unbroken rendering two stable DM candidates that may coexist. There is also a new $U(1)_D$ symmetry with only the added scalar sector being charged under it. This symmetry is broken, so that there is a new massive gauge boson. The gauge Lagrangian associated to the two $U(1)$ symmetries \cite{Pospelov:2007mp,Pospelov:2008zw}, the SM and dark symmetry is:
\begin{align}
	\mathcal{L}_{\rm gauge}=-\frac{1}{4}F_{\mu \nu}F^{\mu \nu}-\frac{1}{4}F'_{\mu \nu}F'^{\mu \nu} - \frac{\varepsilon}{2}F_{\mu \nu}F'^{\mu \nu}~,
	\label{eq:gaugeL}
\end{align}
where, $F_{\mu \nu}$ and $F'_{\mu \nu}$ are the field strength tensors associated to the $U(1)_Y$ and $U(1)_D$ symmetries. The last term  of Eq.~\eqref{eq:gaugeL} implies that there is a mixing between the gauge bosons associated to the two symmetries which after diagonalization results in a massless gauge boson, the photon $\gamma$, and a massive one, the dark photon $\gamma'$ \cite{Pospelov:2007mp,Pospelov:2008zw}. The dark photon acquires its mass once a new complex singlet scalar field develops a vacuum expectation value (v.e.v). A similar scenario was studied in Ref. \cite{Cirelli:2016rnw,Lee:2015gsa,Choi:2017zww}. The presence of this new scalar field does not affect the phenomenology of the model due to the choice of its v.e.v and the chosen parameters presened in  \ref{DMrelic}, and thus we do not discuss it any further. On the other hand, it has been shown that it is possible for this kinetic mixing to vanish \cite{Appelquist:2002mw}, nevertheless, this term will not vanish when a transformation of the gauge fields involves an off-diagonal element, as is considered in this work and previous works such as \cite{Babu:1997st}. Moreover, it has been proposed that the phenomenology associated with this kinetic mixing could be explored at neutrino experiments \cite{DUNE:2020ypp,DUNE:2021tad}, or high- intensity proton beam experiments.

The scalar potential of the added sector which sets the interaction of the new complex singlets $\phi_1$ , $\phi_2$ and the SM Higgs doublet, $H$ reads:
\begin{align}
	& V(H,\phi_1,\phi_2)=  \mu_{\phi_1}^2 |\phi_1|^2 + \mu_{\phi_2}^2 |\phi_2|^2 + \lambda_{\phi_1} |\phi_1|^4 + \lambda_{\phi_2} |\phi_2|^4~+  \\ \nn
	& \lambda_{\phi_1 H}  \ |\phi_1|^2 (H^{\dagger} H)+ \lambda_{\phi_2 H}  \ |\phi_2|^2 (H^{\dagger} H) + \lambda_{\phi_1 \phi_2}  \ |\phi_1|^2 |\phi_2|^2~,
	\label{scalar_potential}
\end{align}
with the last term of the potential implying that the two added scalars may interact with each other. It is important to note that in this potential, both the complex and imaginary part of each $\phi_i$ field have the same mass. This is due to the imposed symmetries. As a result, the DM candidates are the $\phi_i$ fields. Similar scenarios have been studied in Ref. \cite{McDonald:1993ex, Barger:2008jx}. On the other hand,  each new scalar field may interact with SM particles as well as the other scalar field via the dark photon portal. As a result, this portal will be fundamental in setting the phenomenology of the model. Subsequently the new parameters that will have an important effect in the DM phenomenology are the dark fine structure constant $\alpha_D$, the kinetic mixing $\varepsilon$, and the new particles $\phi_1$, $\phi_2$, and $\gamma'$.

In order to explore the phenomenology, the model was implemented in~\texttt{SARAH}~\cite{Staub:2013tta} based on the model developed in Ref. \cite{Sierra:2015fma}.  
 The \texttt{SARAH} package allows for calculations of the relevant interactions among particles, performs anomaly checks and generates output to other packages such as~\texttt{SPHENO}~\cite{Porod:2003um,Porod:2011nf}, which calculates the mass spectrum and other relevant parameters.~\texttt{SARAH} also provides a link to other packages like \texttt{MicrOMEGAS}~\cite{Belanger:2018mqt} which calculates DM observables (relic abundance, spin independent (SI), and spin dependent (SD) cross sections), \texttt{MadGraph}~\cite{Alwall:2014hca} including its plugin \texttt{MadDump}~\cite{Buonocore:2018xjk} which allows for a simulation of DM interaction with electrons at DUNE and SHIP. For the case of DUNE, we include on-axis and DUNE-PRISM analyses. On the other hand, to simulate the production of DM particles at proton collisions we used \texttt{PYTHIA} \cite{Sjostrand:2014zea,Sjostrand:2006za}. Furthermore, neutrino-electron interactions are similar to those of DM particles, so they must be included in the analysis as background; for such processes we used the Monte Carlo generator \texttt{NuWro}~\cite{Golan:2012rfa,zhuridov2020monte}. 

\section{Dark Matter Phenomenology}
\label{DMpheno}
Since this model presents two stable DM candidates that interact with the dark photon, it has a rich phenomenology that could be explored or constrained through the measured relic density as well as direct detection (DD) experiments  that are sensitive to low DM masses.

\subsection{Dark matter constraints}
\label{DMconstraints}
Before exploring the allowed parameter space of the model and choosing appropriate benchmark points, one must consider the different restrictions arising from DM observables:
\begin{itemize}
	\item Following the Planck satellite measurement \cite{Aghanim:2018eyx}, we consider models that fulfill the relic abundance as those that satisfy: 
	\begin{align}
	\Omega \ h^2=0.1200 \pm 0.0012~,
	\end{align}
	allowing a 3$\sigma$ departure from the central value.
	
	\item The DM self interaction cross section is constrained, by observations of galaxy cluster collisions, to be 
	\begin{align}
	\dfrac{\sigma}{m_{DM}} \lesssim \text{few} \times {\rm cm}^2/{\rm g}~,
	\end{align} 
	this imposes limits on the allowed value of $\alpha_D$. For the DM masses considered in this work, $\alpha_D \lesssim 0.5$ \cite{Buonocore:2019esg}.

\end{itemize}

\subsection{Dark matter relic density general framework}
\label{DMgeneral}
\begin{figure}[h]
	\centering
	\includegraphics[scale=0.55]{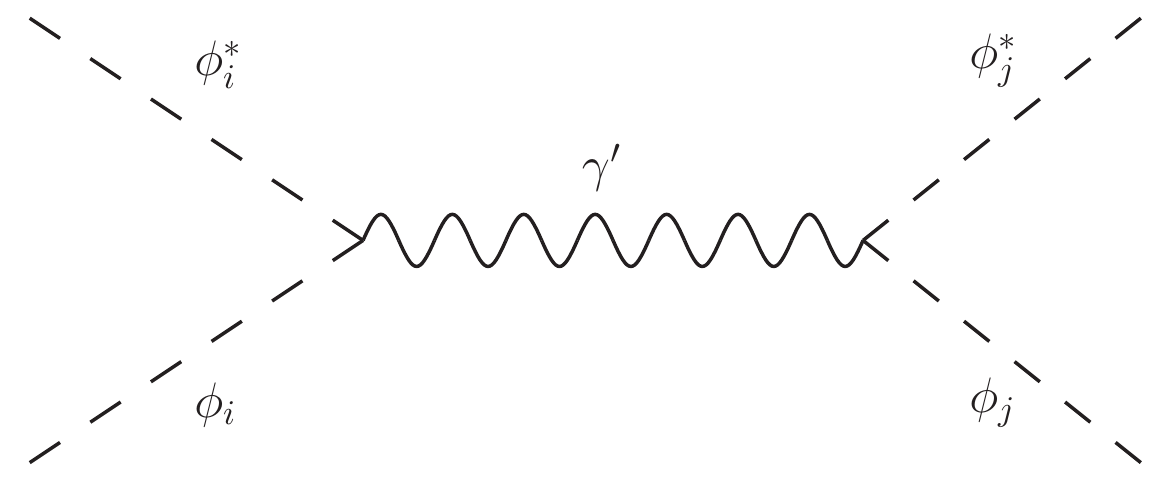}
	\hspace{1.5cm}
	\includegraphics[scale=0.55]{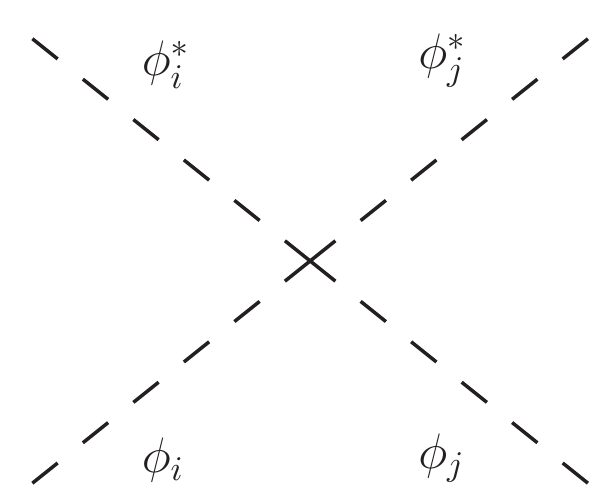}
	\caption{Possible DM conversion channels of the model: Left panel, mediated by the dark photon. Right panel, the four contact interaction.}
	{\label{fig:DM_conv_diagrams}}
\end{figure}

 Because each of these DM candidates, $\phi_1$ and $\phi_2$, is a singlet under the SM symmetries, in principle, few or no channels are available for DM annihilation in the early Universe, thus tending to yield an overabundant density. Nevertheless, due to the presence of the new $U(1)_D$ symmetry and the kinetic mixing between $\gamma$ and $\gamma'$, new annihilation channels appear which allows the DM relic density to be at the observed values for certain portions of the parameter space. In principle, the thermally averaged annihilation cross section is set by the process $\phi \phi \rightarrow \gamma' \rightarrow f \overline{f}$, which depends on dark sector parameters such as $m_{\phi_i}, m_{\gamma'}, \varepsilon$, and $\alpha_D$, as stated in Ref. \cite{Izaguirre:2015yja}. Here $f$ denotes a SM fermion, while $\overline{f}$ denotes a SM anti fermion. This annihilation channel is available as long as the mass of the DM is larger than the electron mass which is the case studied here. This scenario has been investigated in many works such as \cite{deNiverville:2018dbu,DeRomeri:2019kic,Bhattacharya:2016ysw}. Nevertheless, for the present model, the relic abundance could depart from that of previous models due to additional annihilation channels, mainly $\phi_i^{\ast} \phi_i \rightarrow \phi_j^{\ast} \phi_j$ where $i,j$ denote either the $Z_2$ and $Z_2'$ charged scalar particle (shown in the left panel of Fig.~\ref{fig:DM_conv_diagrams}) and through $\phi_i \phi_i^{\ast} \phi_j \phi_j^{\ast}$ which is set by the four contact interaction controlled by the coupling $\lambda_{\phi_1 \phi_2}$, (shown in right panel of Fig.~\ref{fig:DM_conv_diagrams}). These are known as the DM conversion channels, and it has been shown that they may have a profound effect on the relic density \cite{Liu:2011aa}.  
 
\subsection{Relic density and dark matter conversion}
\label{DMrelic}
In order to study DM relic abundance, we used \texttt{MicrOMEGAS-5.2.4}~\cite{Belanger:2018mqt} which allows two DM candidates, giving valuable information about the different annihilation channels that are relevant for a certain point of the parameter space. 

\begin{figure}[h]
	\centering
	\includegraphics[scale=0.35]{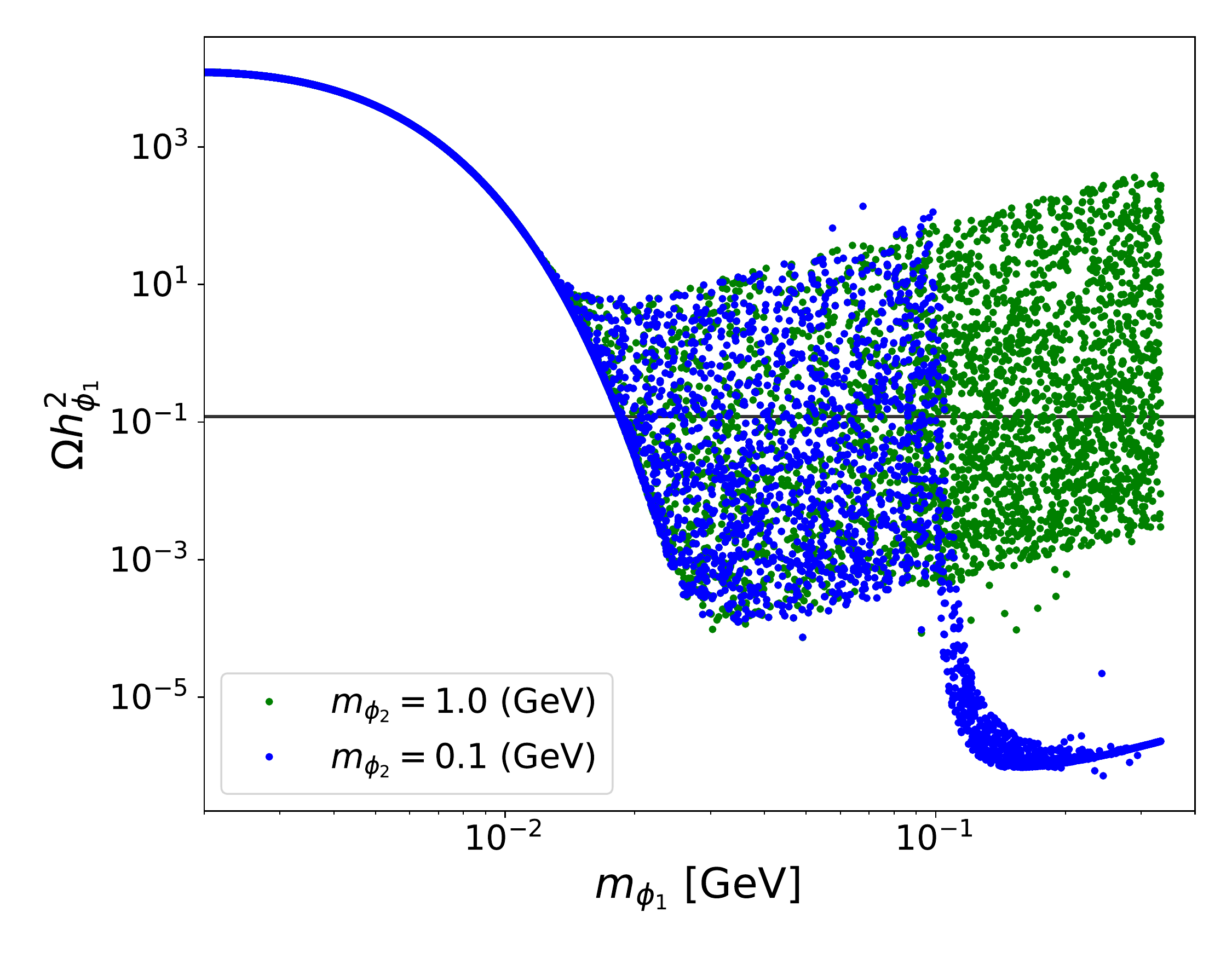}
	\includegraphics[scale=0.35]{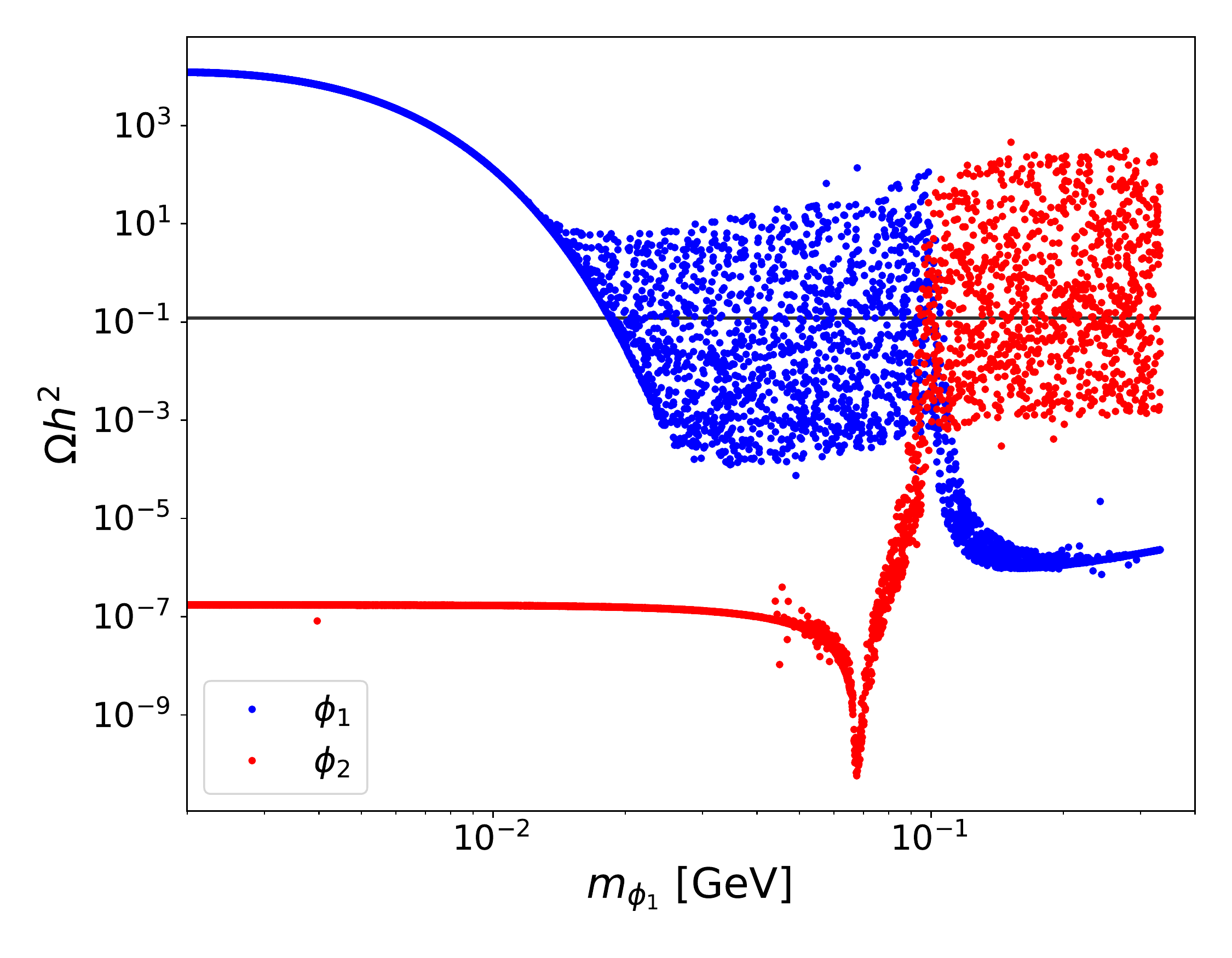}
	\caption{The left panel shows the relic abundance of $\phi_1$ vs. its mass, with $m_{\phi_2}=1.0  (0.1)$ GeV for the green (blue) dots. The right panel shows the relic abundance of $\phi_1$ ($\phi_2$) in blue (red) vs $m_{\phi_1}$ for $\phi_2=0.1$ GeV. For both plots, we set $m_{\gamma'}=3 \ m_{\phi_1}$, while the grey horizontal band corresponds to the measured value of the DM relic abundance.}
	{\label{fig:heavy_light_DM_relic}}
\end{figure}

Because the parameter space is large, we focus on the scenario where $m_{\gamma'}=3 \ m_{\phi_1}$ and $\alpha_D=0.1$, a scenario widely studied in works such as \cite{deNiverville:2018dbu,DeRomeri:2019kic,Breitbach:2021gvv}. This benchmark point allows for a free choice of  $m_{\phi_2}$. Next, we focus on two scenarios for $ m_{\phi_2}$, one where it is fixed and one where $m_{\phi_2}-m_{\phi_1}=$constant. We also set the scalar potential couplings as follows:

\begin{align}
  &\lambda_{\phi_1 H}=\lambda_{\phi_2H}=10^{-10}~,
  \hspace{1cm}\lambda_{\phi_1}=\lambda_{\phi_2}=0.1~,
  \hspace{1cm} \lambda_{\phi_1 \phi_2}=(0-3.0)~.
  \label{scalar_parameters}
\end{align}
Additionally, the coupling of the scalar singlet -- responsible of the $U(1)_D$ symmetry breaking--
to the $\phi_i$ scalar fields  is of the same order than $\lambda_{\phi_i H}$.
For the case of fixed mass of the heavy DM particle, we set $m_{\phi_2}=$ 1.0 and 0.1 GeV. The results of the relic abundance of the light DM particle for this scan are shown in the left panel of Fig. \ref{fig:heavy_light_DM_relic}. As can be seen, for the case of $m_{\phi_2}=$0.1 GeV, once $\phi_1$ becomes heavier than $\phi_2$, its relic abundance becomes suppressed, indicating that this DM particle is being depleted. To further understand this behavior, we present the relic abundance of each DM particle for $m_{\phi_2}=$ 0.1 GeV on the right panel. As is shown, $\Omega h_{\phi_2}^2$ is constant, underabundant, and several orders of magnitude smaller than $\Omega h_{\phi_1}^2$  until $m_{\phi_1}>m_{\phi_2}$. At this point the $\phi_2$ abundance tracks the previous behavior of $\phi_1$ while $\phi_1$ is depleted. This is showing two important things: one, the DM conversion channel is very efficient and proceeds as long as it is viable, two, once the hierarchy of the DM particles mass is inverted, the two particles switch roles. This last point is expected since the two particles have the same behavior under the symmetry groups, thus, the parameter that dominates which DM remains and which gets depleted is the mass ordering. This last point also implies that for $m_{\phi_2}-m_{\phi_1}=$constant we obtain the same results as the case where $m_{\phi_2}=$ 1.0 GeV (same mass ordering). Thus we will not comment any more on this scenario.

\begin{figure}[h]
	\includegraphics[scale=0.35]{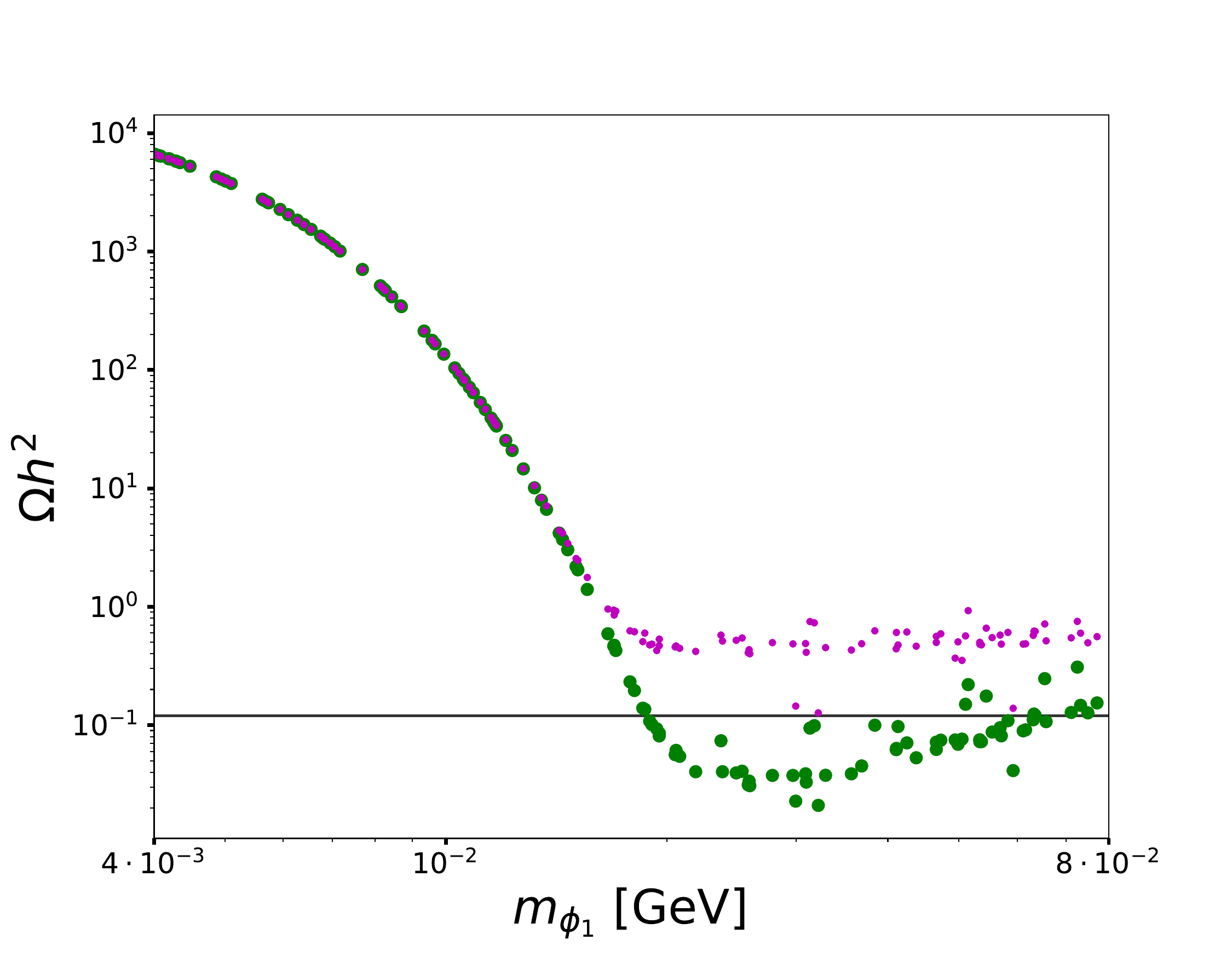}
	\caption{Relic abundance vs. $m_{\phi_1}$ for all channels (green)
          and all channels except the DM conversion channels (magenta) with $\varepsilon^2=10^{-6}$, $m_{\gamma'}=3 \ m_{\phi_1}$ and $m_{\phi_2}=0.1$ GeV.}
	{\label{fig:DM_conv_noconv}}
\end{figure}

Another interesting trend shown both in the left and right panel is that for $m_{\phi_1} \lesssim 0.02$ GeV the relic abundance declines with the DM mass value, yet for $m_{\phi_1}>0.02$ GeV this behavior changes. To explore this, we used the \texttt{MicrOMEGAS-5.2.4} function that allows for exclusion of certain annihilation channels, such as DM conversion. For this, we computed the relic abundance allowing all channels and without the DM conversion channel. The results are presented in Fig.~\ref{fig:DM_conv_noconv} where we set $\varepsilon^2=10^{-6}$ and $m_{\phi_2}=0.1$ GeV. The difference between all allowed channels (green) and channels without DM conversion (magenta) shows is that for $m_{\phi_1}\lesssim 0.02$ GeV, DM conversion channels are not efficient, which means that the relic abundance is set by the annihilation of two $\phi_1$ into SM particles. However, once $m_{\phi_1} \gtrsim 0.02$ GeV the DM conversion channel becomes available and starts to dominate, leading to a significant suppression of the relic abundance compared to the case where there is no conversion. Again, this exhibits the importance of DM conversion on the relic density of the model. It is important to add that this channel proceeds only via the exchange of a dark photon (left panel of Fig.~\ref{fig:DM_conv_diagrams}), that is because the other possibility, which is the four contact interaction (right panel of Fig.~\ref{fig:DM_conv_diagrams}), shows no effect on the relic density when varying $\lambda_{\phi_1\phi_2}$.

\subsection{DM direct detection}
\label{DMDD}

An interesting way to probe DM models is by looking for the evidence of the recoil of a DM particle with a detector located on Earth, this is known as DM direct detection (DD). Due to the interactions of the DM with the dark photon, it is possible for it to recoil against a nucleon via a spin independent (SI) interaction. The CRESST-III experiment \cite{Abdelhameed:2019hmk} has the potential to probe a portion of the DM masses considered here, thus we study if the parameter space is constrained by the experiment. To do so, we again use \texttt{MicrOMEGAS-5.2.4}~\cite{Belanger:2018mqt} which allows for the calculation of the DM-nucleon SI cross section. Nevertheless, we can not directly use the experiment's constraints on the SI cross section due to the following considerations that must be taken into account here:

\begin{enumerate}
	
	\item Due to the presence of two DM candidates, we must rescale CRESST-III restrictions because DD experiments place constraints assuming that one DM particle saturates the relic abundance which has implications on its density in the Earth and then on the expected event rate. For two DM particles the density in the Earth is smaller than for one candidate, thus, we must rescale the cross section in order to impose the restrictions. This is done as follows \cite{Cao:2007fy}: 
	\begin{align}
		\frac{\delta_{\phi_1^0}}{m_{\phi_1^0} }  \sigma_{\phi_1^0}+ \frac{\delta_{\phi_2^0}}{m_{\phi_2^0}}   \sigma_{\phi_2^0}< \frac{\sigma_{0}}{m_0}~,  
		\label{eq:ddrescaling1}
	\end{align}
	where $\delta_{\phi_1^0} (\delta_{\phi_2^0})$ is the fraction of the $\phi_1^0 (\phi_2^0)$, $ \sigma_{\phi_1^0} ( \sigma_{\phi_2^0})$ is the scattering cross section of the $\phi_1^0 (\phi_2^0)$ particle with a nucleon. The parameters associated to the experiment's sensitivity are $\sigma_0$ and $m_0$ which are related to the measured experimental event rate $R_{exp} \sim \rho_0 \ \sigma_0 m_0$  .
		
      \item DD experiments also impose restriction on the DM-nucleon scattering cross section assuming that DM couples equally to protons and neutrons. Nevertheless, this is not always the case, a DM model could show isospin violation. For the present model, we find that DM particles exhibit such violation. This, again, implies that the cross-section must be rescaled, and works such as \cite{Feng:2011vu,Yaguna:2016bga} have found that such procedure must be done as follows:
	 	\begin{align}
	 	  \dfrac{\sigma_p}{\sigma_N^Z}= \dfrac{ \displaystyle \sum_i \
                    \eta_i \mu_{A_i}^2 \ A_i^2}{\displaystyle \sum_i \
                    \eta_i \mu_{A_i}^2[Z+\ (A_i^2-Z) f_n/f_p ]^2 }~.
	 	\label{eq:ddrescaling2}
		\end{align}
 	Here, $\sigma_p$ and $\sigma_N^Z$ are the DM-nucleus and DM-nucleon SI cross section, respectively. This equation takes into account that the DM detector could be composed of different isotopes $A_i$ with $\eta_i$ being the fractional abundance of the isotope, $\mu_{A_i}$ is the DM-nucleus reduced mass, $Z$ is the number of protons, and, $f_p$ ($f_n$) is the DM-proton (-neutron) coupling.
 	Due to isospin violation, the experiment's sensitivity is reduced, which implies that the restrictions on the cross section must be rescaled by the factor $\sigma_p/\sigma_N^Z$.
 	
 	Taking into account both Eqs.~\eqref{eq:ddrescaling1} and~\eqref{eq:ddrescaling2}, we find that the CRESST-III results do not impose constraints on the parameter space of the model. Thus, for the time being, this model can not be explored by DD experiments. 
 	
 \end{enumerate}
 
 \begin{figure}[h]
	\includegraphics[scale=0.45]{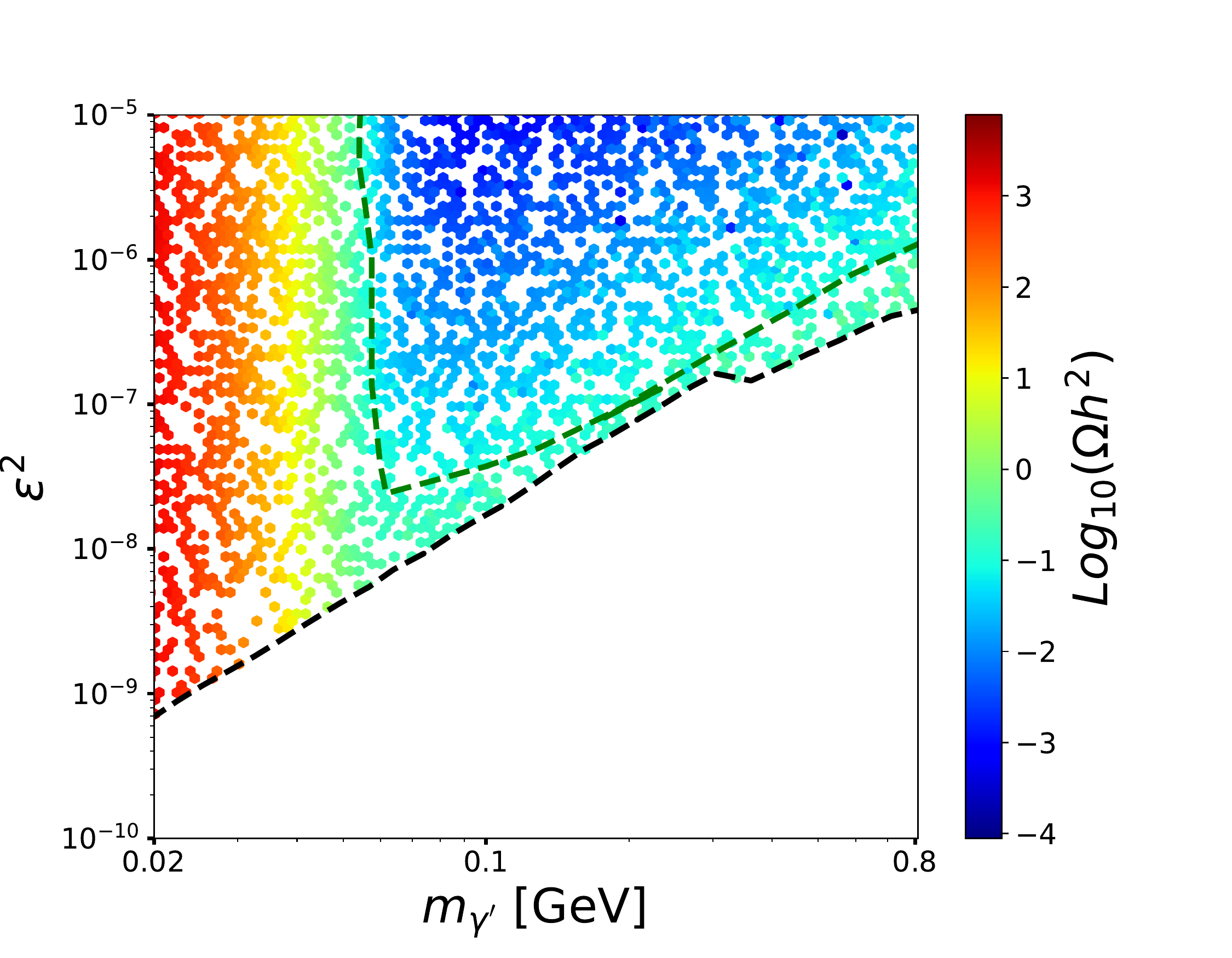}
	\caption{Kinetic mixing vs. the dark photon mass where the color code denotes the relic density. The green (black) dashed line represents the points where the measured relic abundance is obtained for a two (one) DM component model.}
	\label{dmcolor}
\end{figure}

Since this model may not be currently probed through DD experiments, we consider other possibilities such as high-intensity proton beam experiments, e.g. DUNE \cite{Abi:2020wmh} and SHiP \cite{Shirobokov:2021bno}. These experiments have the capacity to explore sub-GeV DM models, and a discussion on this  is found in Sec.~\ref{sec:dm@nu-facilities}. These experiment's sensitivity depends on $\alpha_D$, $\varepsilon$, and the dark sector masses. For the aforementioned scenario ($\alpha_D=0.1$, $m_{\gamma'}=3 m_{\phi_1}$, and $m_{\phi_2}=1.0$ GeV), the relevant parameters for both the relic abundance and the sensitivity become then $\varepsilon$ and $m_{\gamma'}$. Thus, we present the results of $\Omega h^2$ (color map) in the plane of the two relevant parameters in Fig.~\ref{dmcolor}. Since we want to compare to previous works such as \cite{DeRomeri:2019kic}, we also present the measured relic abundance for one component DM with the black dashed line and for the present model with the green dashed line. The plot shows that the two contours differ greatly, while the one component model is nearly diagonal, the two component has a triangular shape in the $\varepsilon^2$-$m_{\gamma'}$ plane. Here, we want to emphasize that even for the two component model, only one DM particle remains today, due to DM conversion. Nevertheless, the two DM particle model is very different to the one DM particle model, this in turn changes the portion of the parameter space where relic density is saturated.

\section{Dark Matter at high-intensity proton beam facilities}
\label{sec:dm@nu-facilities}

\begin{figure}[t]
	\centering
	\includegraphics[scale=0.7]{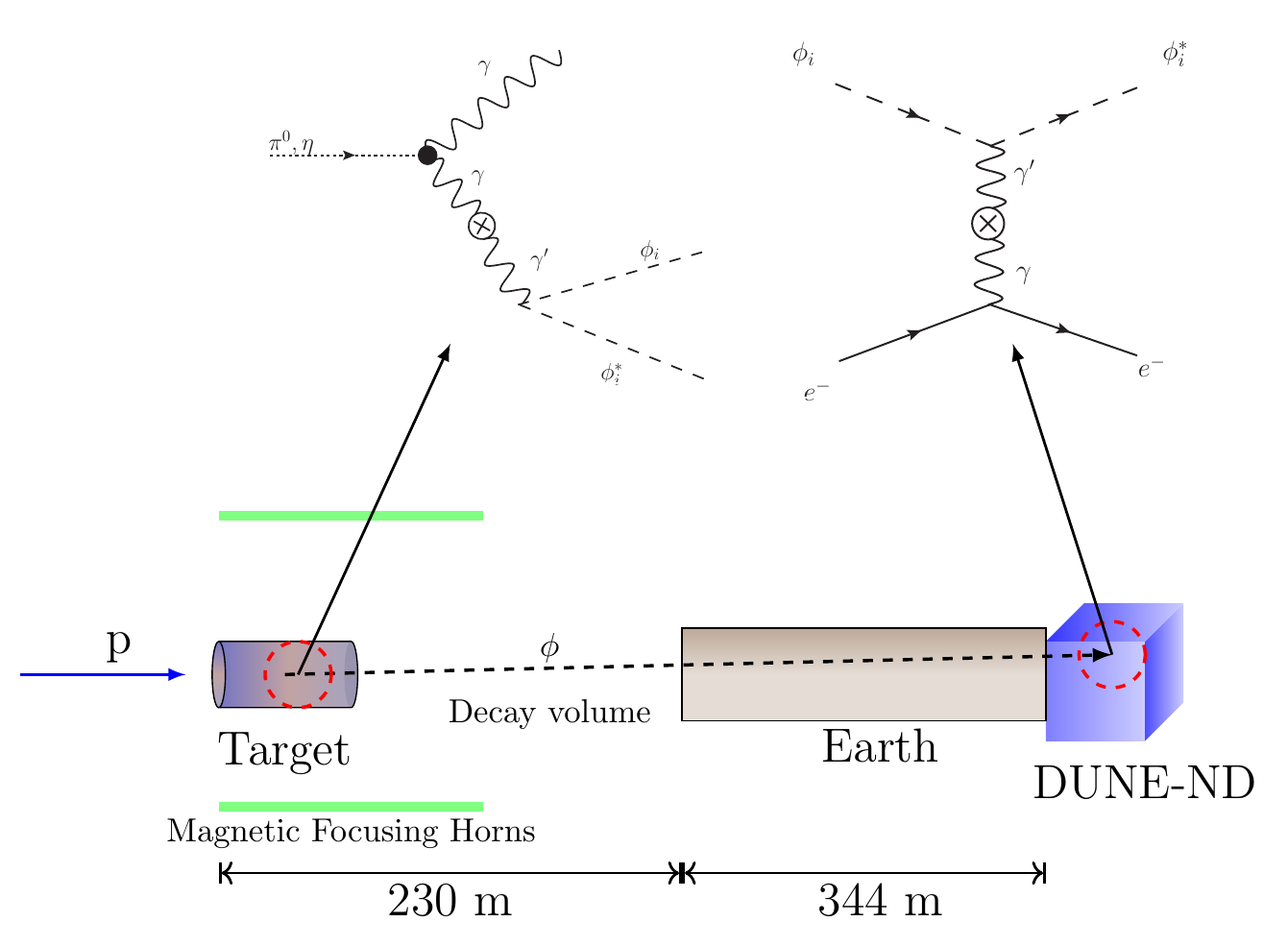}
	\caption{Schematic illustration of
          DM searches at DUNE ND. A high intensity
          proton beam with an average energy of
          $120$~GeV hits a fixed target
          made of carbon. After the collision,
          mesons are created and later
          decay to DM species which travel $574$~m
          and interact with electrons in the DUNE ND.
        }
	{\label{fig:schematic_DUNE_ND}}
\end{figure}

High-intensity proton beam experiments provide a great opportunity to explore sub-GeV dark sectors. After the proton beam collision with the target, a large number of mesons are generated (such as $\pi^0$ and $\eta$) with a significant fraction of them decaying to two photons. A portion of those photons will convert into the dark photon thanks to the kinetic mixing. The dark photon then decays to two DM particles, and these particles will travel through Earth to the detector nearly unperturbed due to their weakly interacting nature. Once in the detector, they may interact with particles like electrons via the exchange of a photon-dark photon, see Fig.~\ref{fig:schematic_DUNE_ND}. Thus, to simulate the process, one must include the meson production and decay, DM production and interaction with the detector, as well as the neutrino background. This section describes the steps to simulate the aforementioned processes, in addition to the results on neutrino backgrounds.

\subsection{Fluxes}
\label{sec:fluxes}
In high-intensity experiments, proton beams impact targets of diverse compositions (graphite and lead), resulting in the production of different particles. For instance, the collision produces neutral mesons (e.g., $\pi^{0}$ and $\eta$) with rates depending on beam energy and target composition. In our case, the flux of these mesons was simulated with the help of the Monte Carlo generator  
{\tt PYTHIA 8.230}~\cite{Sjostrand:2006za,Sjostrand:2014zea}.
In the case of relatively low-energy fixed
target experiments, the flag {\tt SoftQCD:all} is recommended
to use~\cite{Berryman:2019dme}.
 Here, we consider the proton-carbon collision and hence the flux production of $\pi^{0}$ and $\eta$; studying the dominants decays of $\pi^{0} \to \gamma \gamma$ (BR $\approx$ 98$\%$), $\eta\to \gamma \gamma$ (BR $\approx$ 39$\%$), and $\eta \to 3\pi^{0}$ (BR $\approx$ 33$\%$) \cite{ParticleDataGroup:2020ssz}. The table~\ref{table:neutral_meson_production} displays the  results obtained in the simulation of neutral-meson events normalized on the number of protons on target ($N_{\text{POT}}$).

\begin{table}[htp]
\centering
\begin{tabular}{|l|c|c|}
\hline
\multicolumn{3}{|c|}{\texttt{SoftQCD:all}}\\
\hline \hline
     Process&$p \text{C} \to \pi^{0}\mathcal{X}$& $p \text{C} \to \eta \mathcal{X}$  \\
\hline
      N of part./$N_{\text{POT}}$&4.1& 0.50  \\
\hline         
\end{tabular}
\caption{Number of neutral mesons  normalized to $N_{\text{POT}}$ produced in the collision proton-Carbon ($E_{\text{beam}}=120$ GeV), simulated with \texttt{PYTHIA} and using the flag {\tt SoftQCD:all}. Here $\mathcal{X}$ refers to other particles produced during the collision.}
\label{table:neutral_meson_production}
\end{table}

Neutral mesons decay just after their creation into photon pairs. These photons, via the kinetic mixing mechanism, can give rise to the flux of dark photons via the process $X \to \gamma \gamma^{\prime}$ (being $X=\pi^{0}$ or $\eta$) and the subsequent decay of  $\gamma^{\prime} \to \phi \phi{^*}$ \footnote{We do not consider here
the production of DM via proton bremsstrahlung
$pp \to pp \gamma^{\prime}$ and the production
process induced by leptonic secondary
particles and their bremsstrahlung. In the kinematic
region under consideration, these processes,
if not negligible, are subdominant as it was shown in Ref.~\cite{Celentano:2020vtu}.}. This reaction can proceed via on-shell
or off-shell $\gamma^{\prime}$. Here, we consider
the $\gamma^{\prime}$ decay is dominated by the on-shell mode. In that case, the branching ratio (under the narrow-width approximation) to the new channel fulfills~\cite{deNiverville:2011it, Gardner:2015wea}

\begin{eqnarray}{\label{eq:reaction}}
  \text{Br}(X \to \gamma \phi \phi^{*} ) \simeq \text{Br}(X \to \gamma \gamma)\times 2
  \varepsilon^{2}\Bigg( 1-\dfrac{m_{\gamma^{\prime}}^{2}}{m_{X}^{2}}\Bigg )^{3} \text{Br}(\gamma^{'} \to \phi \phi^{*}) ~.
\end{eqnarray}

With $X=\pi^{0}$ or $\eta$. Once produced, the DM candidates travel up to ND and interact with the electrons
of the liquid argon via $\phi e^{-} \to \phi^{*} e^{-}$ scattering.

Besides, charged mesons like pions and $K$-mesons, among others, are produced in the primary collision of the proton beam and target. In the case of DUNE, these charged mesons are collimated and selected by a system of magnetic horns. Depending on the orientation of the magnetic field of the horn, antimuons (muons) and two flavors of neutrino (antineutrino) survive after the decay of charged pions. Also, $K^{+}(K^{-})$-mesons, among other particles, are selected by the horn. The muons and $K$-mesons decay to form the  $\nu_{\mu}, \nu_{e}$ or $\bar{\nu_{\mu}}, \bar{\nu_{e}}$ fluxes. 
These fluxes are simulated by the collaboration DUNE, taking into account a series of effects, such as the interaction of the particles with the magnetic horn or the rock surrounding the beam  \cite{LFFluxesDUNE}. The uncertainty in the simulation of the neutrino and antineutrino fluxes plays a crucial role in the physical program of LBNF, in particular, in oscillation physics \cite{CDRND2021}. 
After their production, the  $\nu_{\mu}, \nu_{e}$ or $\bar{\nu_{\mu}}, \bar{\nu_{e}}$ fluxes travel to the ND, where they could interact with the electrons and nuclei of the liquid argon.

\subsection{Signal and Background at the DUNE Near Detector}
\label{sec:signalbackground}

It is relevant to find an approach, based on events distributions, to distinguish the recoil of electrons in DUNE ND produced by the DM candidates and neutrinos. 
The detector registers the recoil of electrons
that ionize and scintillate the active material
of the detector producing a signal that is recorded.
Because of the forward-going nature of the
electron track, neutrinos of the beam will
produce events that look similar to DM
signal through $\nu e^- \to \nu e^-$
scattering ($\nu-e$), which is purely
electroweak process between neutrinos and atomic
shell electrons\footnote{ $\nu-e$ is not
affected by hadrons or nuclear dynamics
producing a clean signal of a very forward
electron on the detector.}.
Electrons can also appear via $\nu_{e}$ induced
Charged-Current Quasi-Elastic ($\nu_{e}$-CCQE)
scattering or $\nu_e\hspace{0.08cm} ^{40\hspace{-0.05cm}}\text{Ar} \to e^- \mathcal{N}$,
in which the final nucleon activity ($\mathcal{N}$)
is missed. Since the DUNE beam will
operate in two horn currents, focusing positive
and negative mesons producing a beam mostly composed of
neutrinos and antineutrinos, respectively;
each mode will drive out to different rates
for the $\nu_{e}$-CCQE and $\nu-e$ backgrounds.

To see the influence of these standard neutrino channels in the DM searches, we evaluate $\nu-e$ and $\nu_{e}$-CCQE backgrounds using the \texttt{NuWro} event generator~\cite{Golan:2012rfa}. \texttt{NuWro} implements all important interactions in CC (charged currents) and NC (neutral currents) neutrino-nucleus processes and neutrino-electron scattering (channel added to the general module~\cite{zhuridov2020monte}). It is relevant to point out that although the $\nu_{e}$ flux at DUNE ND is $\sim$1$\%$ of the total neutrino flux (in the on axis-position), the $\nu_{e}$-CCQE cross-section is over three orders of magnitude larger than the one for $\nu-e$ scattering \cite{CDRND2021}. Nevertheless, the $\nu_e$-CCQE background can be reduced by choosing appropriate kinematical cuts; the electron recoil energy $E_{e}$ and the electron angle with $z$ axis $\theta_{e}$, particularly the composed function $E_e\theta_e^2$. The $\nu-e$ scattering is constrained to $E_{e}\theta_{e}^{2}\leq 2m_{e}$ for a very forward electron along the direction of the neutrino beam (with $E_{e}\ll E_{\nu}$ in the maximum value).

Since DM$-e$ scattering shares the same kinematical behavior of $\nu-e$ with an identical topology in the Feynman diagrams, and hence the same cut $E_{e}\theta_{e}^{2}\leq 2m_{e}$, the signal has a similar signature regarding $\nu-e$ distribution. This fact converts $\nu-e$ into an irreducible background according to signal spectrum in the $E_{e}\theta^2_{e}$ quantity. The $\nu_{e}$-CCQE has additional nuclear effects modifying the very forward nature for the outgoing electron, and therefore $E_{e}\theta_{e}^{2}$ spectrum looks different to the signal, making it a reducible background \cite{DeRomeri:2019kic}. DUNE ND will have the capability to reconstruct the electron energy with a resolution of 5$\%$ and the angle with a resolution of $12$ mrad ($E_{e}>2$ GeV)~\cite{CDRND2021}. With these values for DUNE ND, it is possible to compute the expected acceptance on $E_{e}\theta_{e}^{2}$ giving a maximum of 2 MeV rad$^{2}$. Using this resolution, it will be possible to cut
more than 99$\%$ of events according to the electron energy and angle square in $\nu_{e}$-CCQE background (see appendix \ref{app:events} and table \ref{tab:numode} for details). In what follows, we consider the impact of the $\nu_e$-CCQE-events distribution stored in $E_{e}\theta^{2}_{e}<2$ MeV rad$^{2}$.

\begin{figure}[htp]
\centering
\includegraphics[scale=0.44]{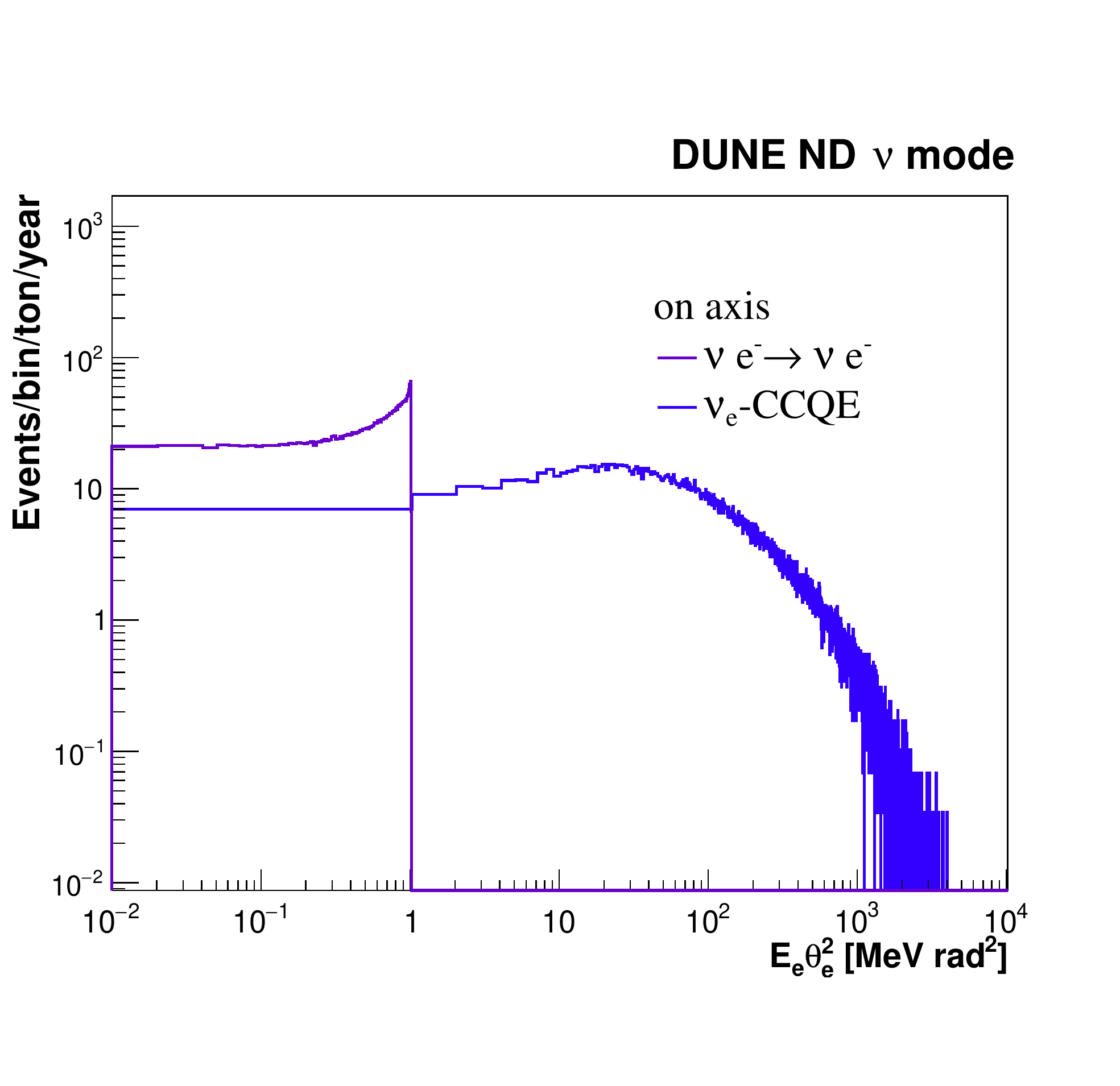}\hspace*{-0.5cm}
\includegraphics[scale=0.44]{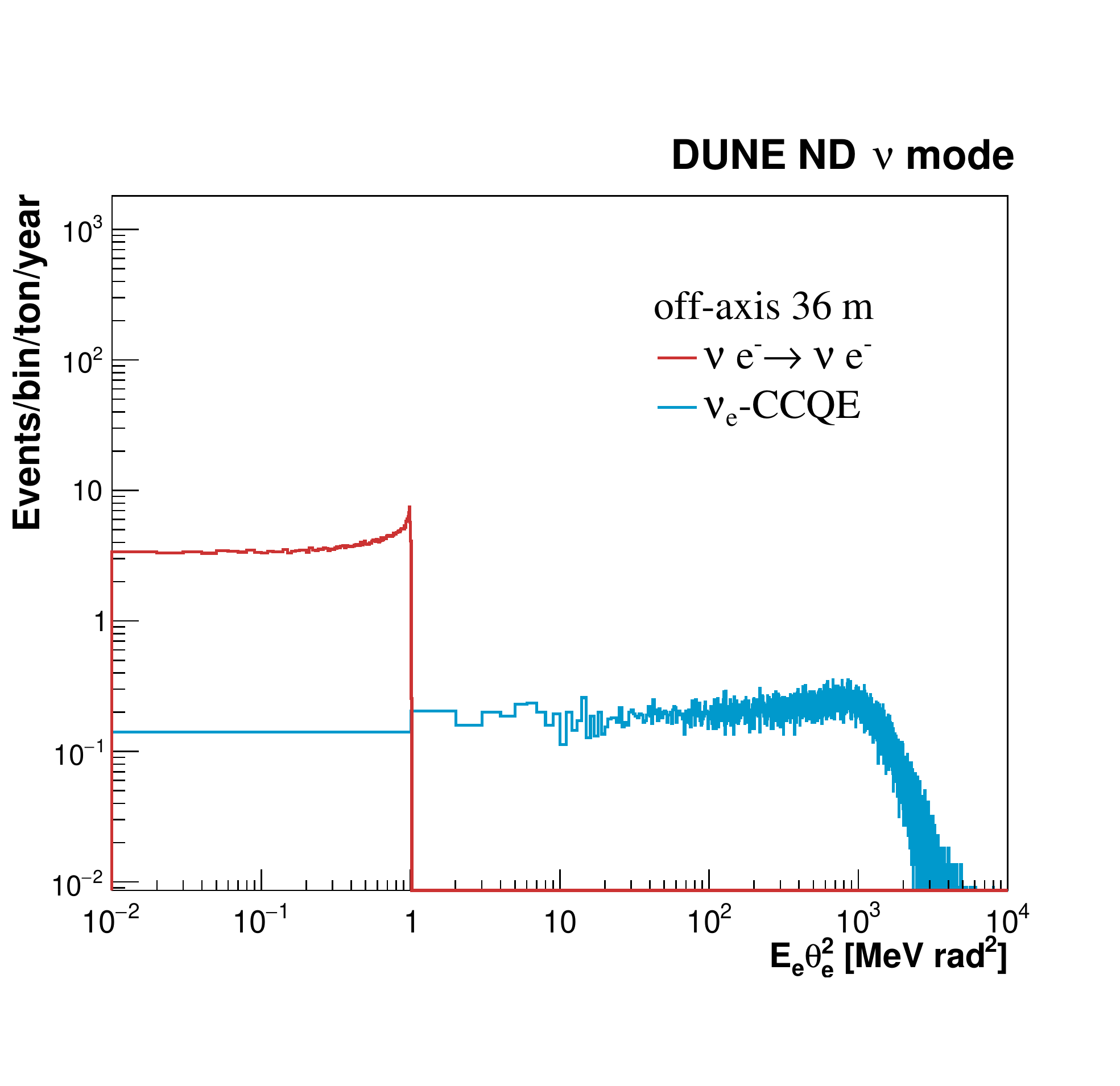}
\vspace*{-1.0cm}
\caption{Neutrino-electron and CCQE event rates as a function of the outgoing electron energy times angle squared for the extremal positions (on-axis and off-axis at 36 m) of the DUNE ND within $\nu$  mode (normalized to total events/year/ton). Bin size for $\nu_e$-CCQE is $1$ MeV rad$^{2}$ and for $\nu-e$ is 0.01 MeV rad$^{2}$. }
\label{fig:Bkg1}
\end{figure}

Fig.~\ref{fig:Bkg1} shows the $E_e\theta_e^2$ distribution of 
$\nu-e$ scattering and $\nu_e$-CCQE events in $\nu$ mode for the on-axis and off-axis at 36 m. For both configurations of the detector, more than 99$\%$ $\nu_e$-CCQE background events have $E_{e}\theta_{e}^{2}>2$ MeV rad$^{2}$. This fact means that we can put a cut over this variable, which accepts the complete $\nu-e$ background, and less than $0.5\%$ the $\nu_{e}$-CCQE background. 

\begin{figure}[htp]
\centering
\includegraphics[scale=0.44]{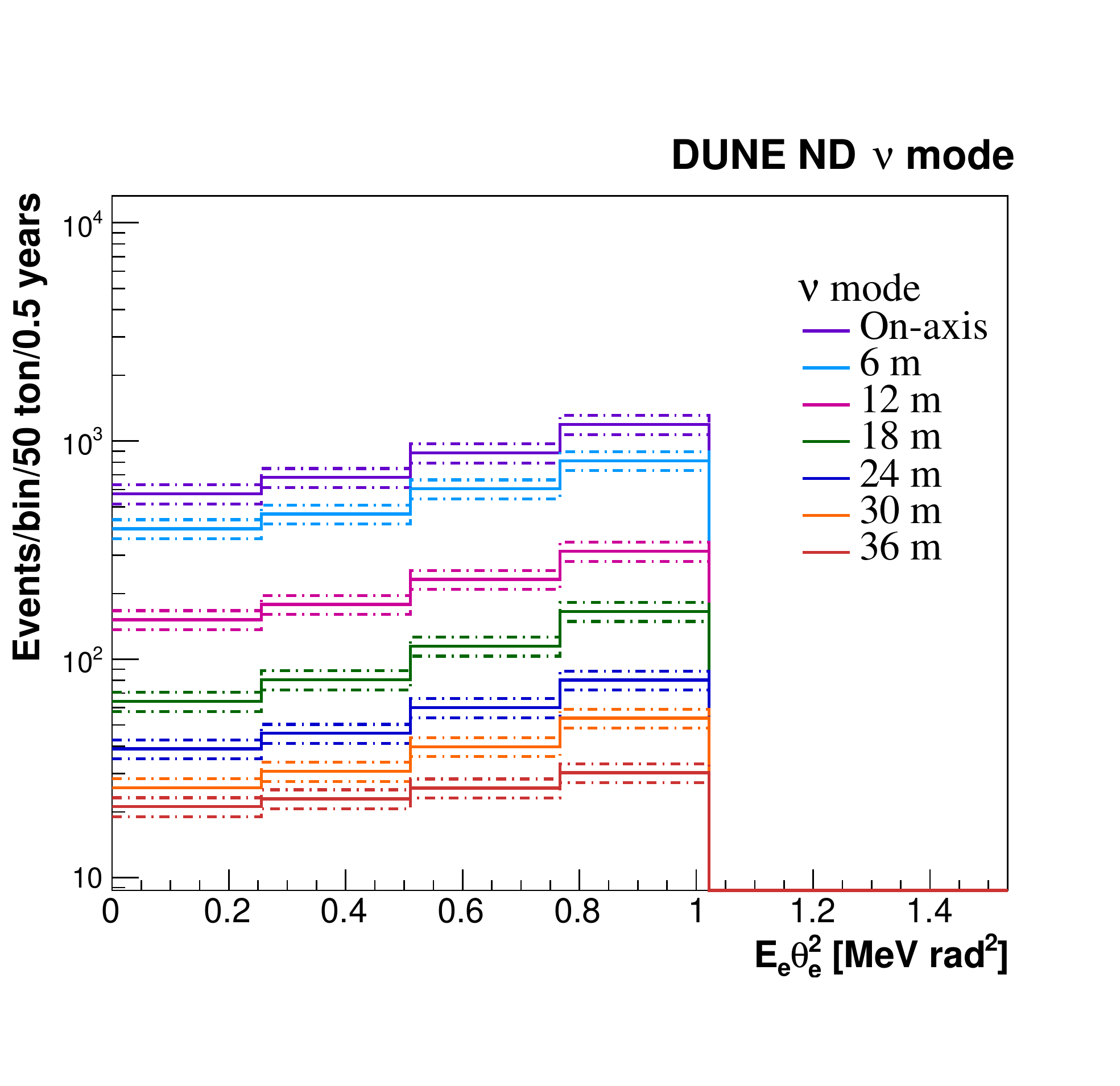}\hspace*{-0.5cm}
\includegraphics[scale=0.44]{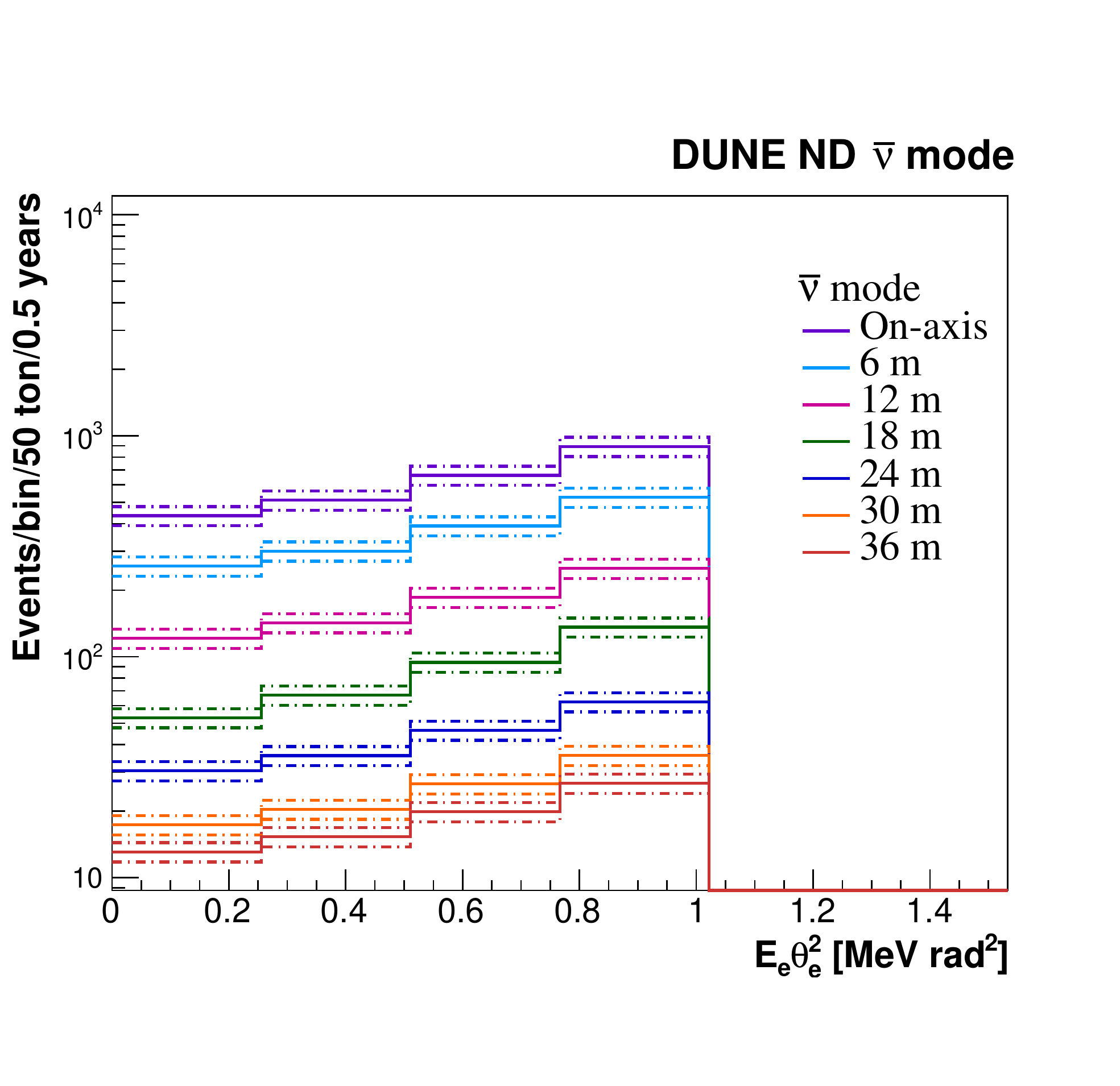}
\vspace*{-1.0cm}
\caption{Neutrino-electron event rates in the DUNE ND considering $\nu$ (left)
  and $\bar{\nu}$ (right) modes as a function of $E_{e}\theta^{2}_{e}$
  for the on-axis and six off-axis positions.
  Distributions allow a 10$\%$ normalization uncertainty for both of them (dashed-point line).}
\label{fig:Bkg2}
\end{figure}

In DUNE-PRISM, the presence of different neutrino flavors in different positions is a powerful tool to study signal-background sensitivity.  Fig.~\ref{fig:Bkg2}  depicts the
$E_{e}\theta_{e}^{2}$ distribution of $\nu-e$ scattering events for six off-axis positions in the $\nu$ and $\bar{\nu}$ modes without any cut of angular or energy resolution. The $E_{e}\theta_{e}^{2}$ spectra show how $\nu-e$ background decreases when the detector is moved across several off-axis positions. Besides, the normalized histograms for $\nu-e$ events distribution are reduced when the beam is running in the anti-neutrino mode.

\begin{figure}[ht]
\begin{center}
\includegraphics[scale=0.44]{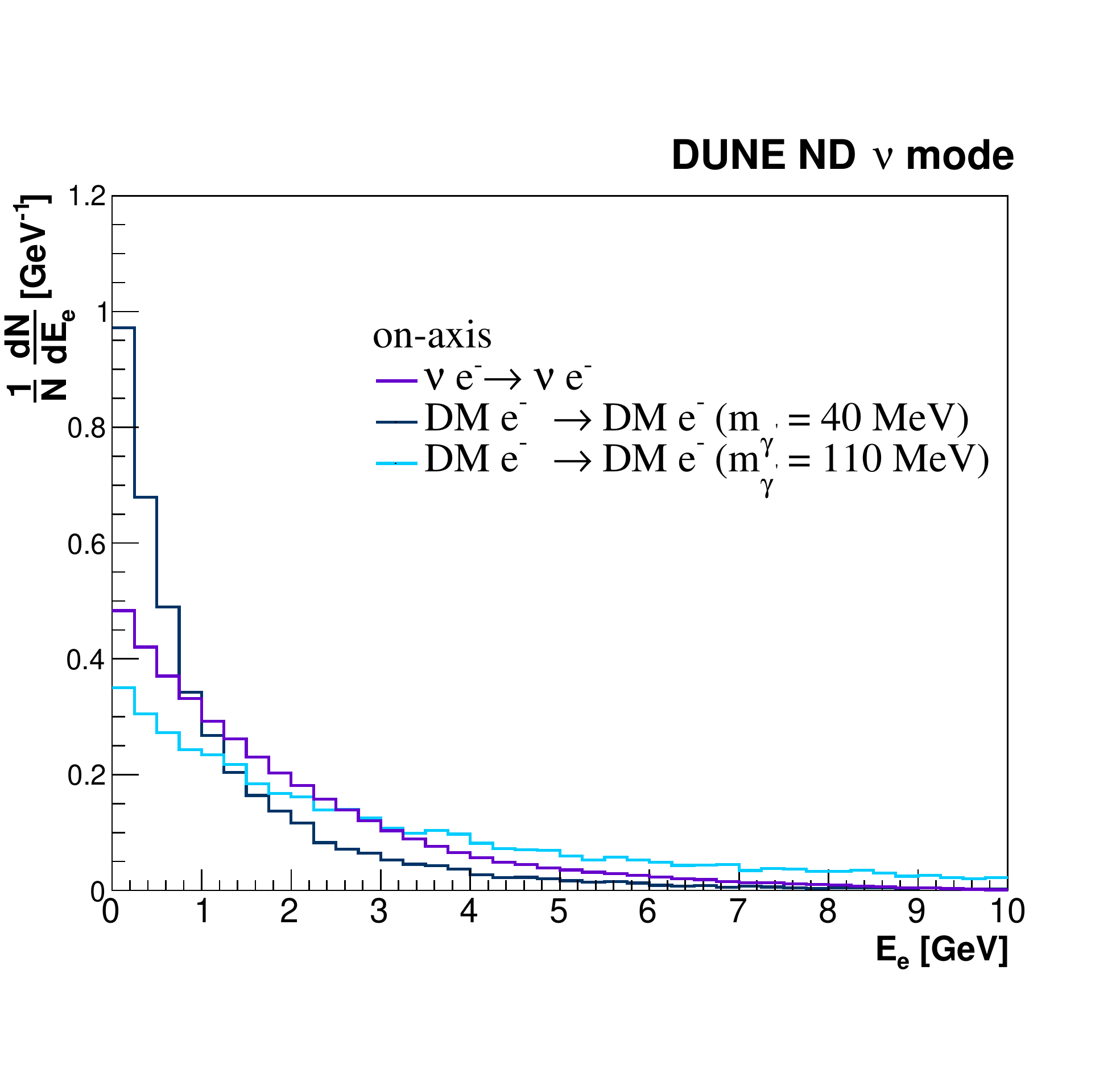}\hspace*{-0.75cm}
\includegraphics[scale=0.44]{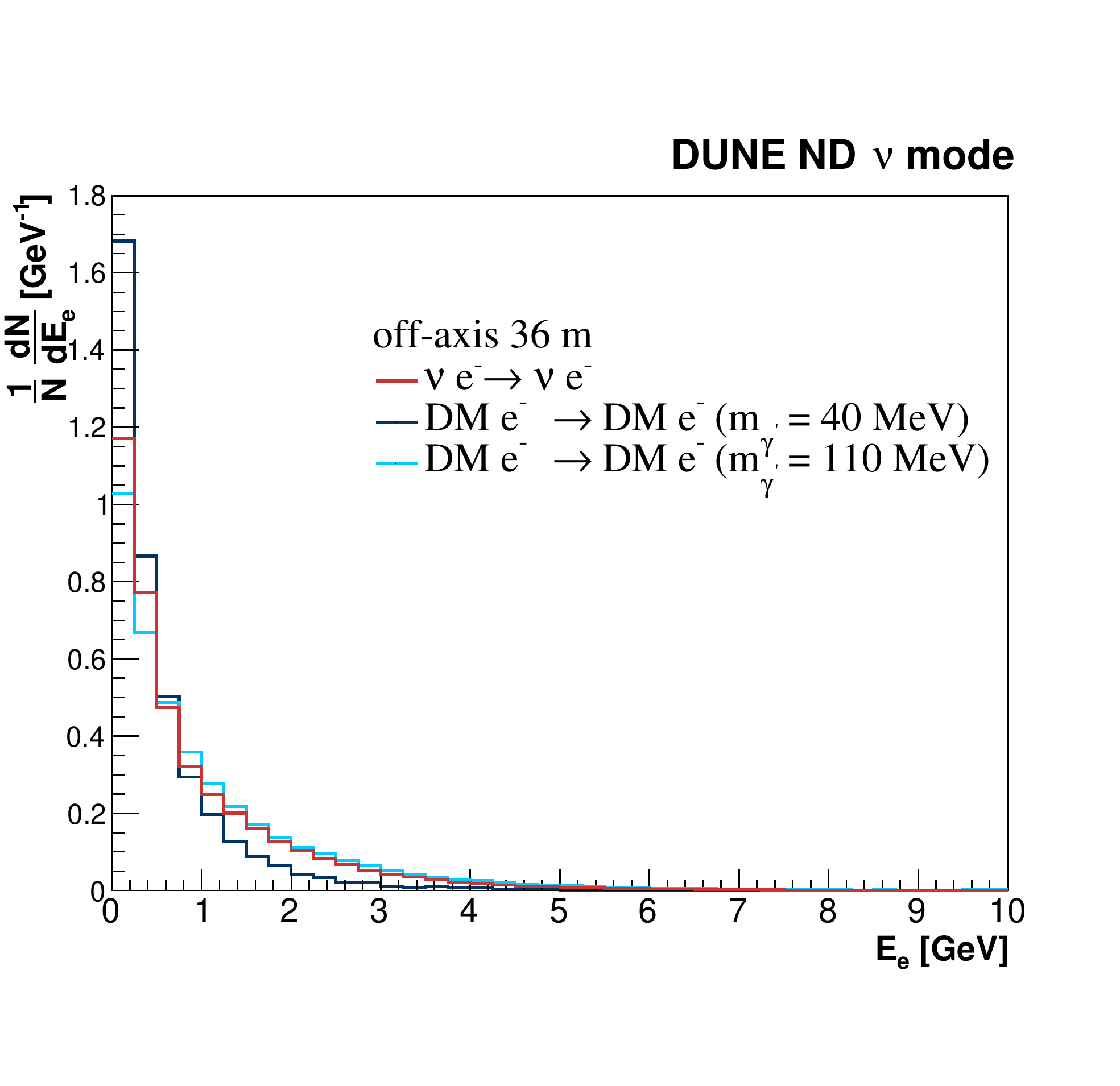}
\vspace*{-1.0cm}
\caption{Electron energy distribution for $\nu-e$ background events (violet) and DM $e^-$ scattering signal with $m_{\gamma^\prime} = 40$ MeV (dark blue) and $m_{\gamma^\prime} = 110$ MeV (cyan) for on-axis and one off-axis configuration at 36 m. In both positions, signals are constrained to have $m_{\phi_{1}}=m_{\gamma'}/3$ and $m_{\phi_{2}}=1$ GeV.}
\label{fig:ElectronEnergyMainText}
\end{center}
\end{figure}

To find out critical discrepancies between signal and $\nu-e$ background, we consider the spectrum of the final electron energy (bins of $0.25$ GeV) in the on-axis and off-axis configurations.  Fig.~\ref{fig:ElectronEnergyMainText} displays $\nu-e$ and $\rm{DM}-e$ signal for two benchmark scenarios in the maximal and minimal positions of the DUNE ND. For both on-axis and off-axis (36 m) positions of the LAr detector, the different background distribution profiles regarding the final electron energy provide an additional strength to search for DM at DUNE ND, since these discrepancies impact the signal-background ratio. The electron energy distribution works efficiently in the identification of reference points where signal and background differ, everything motivating a sensitivity study using this kinematic variable.

\subsection{The impact of constraints on sensitivity}
\label{sec:constraints}

  In addition to the constraints
  coming from DM phenomenology given
  in section~\ref{DMconstraints}, we also
  took into account restrictions
  on the parameter space 
  coming from B-factories and beam-dump experiments.
The most stringent constraints of dark photon comes from BaBar searches.
Light dark photon production
  at BaBar B factory, takes place
  via the reaction $ e^{+}e^{-} \to \gamma \gamma^{\prime}
  \to \gamma + {\rm{invisible}}$, for $m_{\gamma^{\prime}} > 2m_{\rm{DM}}$, or 
$e^{+}e^{-} \to \gamma \gamma^{\prime} \to \gamma e^{+}e^{-}$,
  when $m_{\gamma^{\prime}}<2m_{\rm{DM}}$~\cite{Lees:2014xha, Lees:2017lec}.
  The latter restriction corresponds to the gray region
  shown in Fig.~\ref{plot:sensitivity}.
  On the other hand, beam-dump
  experiments put limits
  on light DM produced via decay
  of neutral mesons which are produced
  in the collision of a proton ( or electron) beam
  against a fixed target. The most stringent
  limit is imposed by NA64~\cite{Banerjee:2019dyo,Andreev:2021xpu},
  which exclusion region is also displayed
  in the lighter gray in Fig.~\ref{plot:sensitivity}.

  \section{Sensitivity analysis}
  \label{sec:sensitivity_analysis}

 The set of experiments used in the sensitivity analysis are briefly described in what follows.
\begin{itemize}

  \item 
    {\bf DUNE}: In the case of DUNE ND, a proton beam at $120$~GeV
    collides against a fixed target made of carbon.
    After the collision, a huge amount of particles are produced,
among them,  mesons, baryons, and other intermediate
    unstable particles.
    These particles
  later decay to photons, neutrinos, and other stable states. After that, the
  photons, neutrinos, and neutral particles,  including possible
  WIMPs travel $574$~m
  until they reach the DUNE ND. The detector will have a fiducial mass of 50 tons of liquid argon  and
  a rectangular parallelepiped shape with dimensions of $6$~m wide,
  $2$~m high and $3$~m deep~\cite{AbedAbud:2021hpb} (see  schematic in
  Fig.~\ref{fig:schematic_DUNE_ND}).
  Inside the detector, the incoming neutrinos,
  as well as possible DM particles, scatter off electrons
    of the argon's atoms, leading to a signature with an
    energetic electron at the final state. The
    distinction between $\nu-e$ scattering events (the background)
    and  DM$-e$ scattering events (the signal) is
    achieved by comparing the electron recoil energy and $E_e \theta_{e}^2$ distributions of both signatures.
  It is expected that over seven years of running, a total of $7.7\times10^{21}$ $N_{\text{POT}}$ will be delivered.
  The ND will be able to take data at different off-axis  positions.
  Among other things, this will improve the potential exclusion
  reach of DUNE on the model because it increases the ratio between signal to background.
  On the other hand, the combination
  of all the results, those  obtained for the on-axis  and
   those for the off-axis configurations  allows to 
   explore a bigger parameter space
   in comparison
    with the use of the on-axis or off-axis data independently.

\item  
  {\bf SHiP}:  Regarding this experiment, a proton beam  at $400$~GeV hits a fixed target made of
  a thick heavy-metal hybrid~\cite{SHiP:2020noy}. In the collision, neutral stable
  particles are produced that then travel $38$ m until reaching the Scattering Neutrino Detector (SND),
  where $\nu-e$ scattering and  DM$-e$ scattering take place. In five years
  of running a total of $2 \times 10 ^{20}$ $N_{\rm{POT}}$ will be delivered.

 \item  
   {\bf NO$\nu$A}: In this case  a proton beam  at $120$~GeV strikes a $1.2$ m graphite target.
   After the collision takes place, neutral stable
   particles (neutrino, DM species) are produced
   and then travel $990$ m until reach NO$\nu$A ND
   located $14.6$ mrad away from the central axis of the neutrino beam.
   We used a $2.97 \times 10^{20}$ $N_{\rm{POT}}$ of data taking~\cite{Bian:2017axs,Acero:2019qcr,Acero:2019ksn}
   for the recasting analysis.
  
\end{itemize}

 \begin{figure}[h!]
	\includegraphics[scale=0.45]{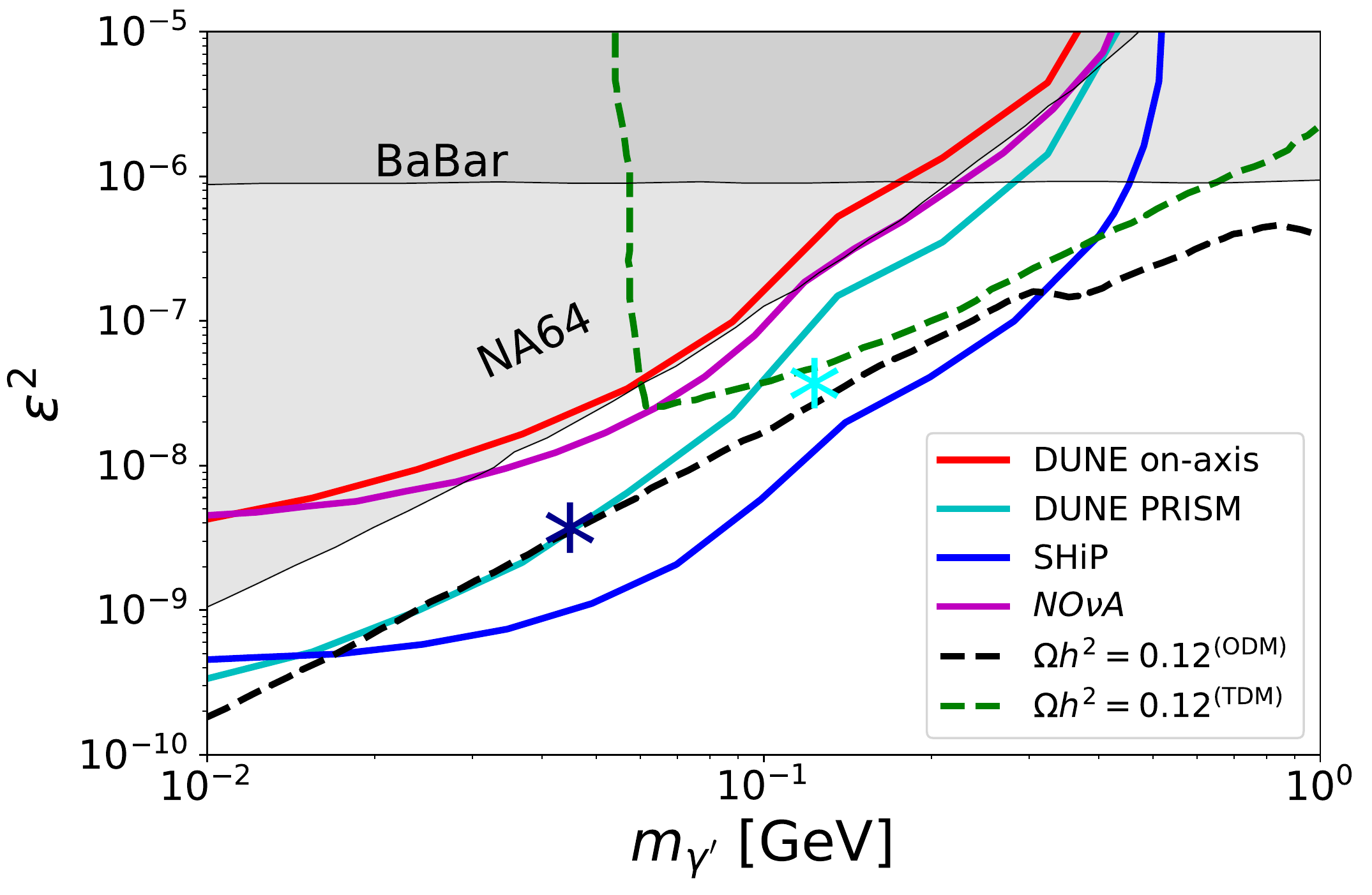}
	\caption{ Projected sensitivity at $90$\%~CL. The
           red, cyan and blue solid line stand for
          the expected sensitivity of DUNE on-axis, DUNE-PRISM and SHiP respectively.
          The recasting of No$\nu$A is displayed  in the magenta line and 
          current constraints are given in the  gray regions.
          The benchmark scenario considered in the
          analysis correspond to $\alpha_D = 0.1$ , $m_{\gamma^{\prime}} = 3m_{\phi_1} $
          and $m_{\phi_2}=1.0$~GeV.
        }
	\label{plot:sensitivity}
\end{figure}

In Fig.~\ref{plot:sensitivity}, the results
on the sensitivity and current exclusion limits are displayed as a function
of the dark photon mass. The gray regions 
correspond to BABAR and NA64 experimental limits at $90$\%~CL.
The magenta line represents the current bound from NO$\nu$A, the region
above that line is excluded at $90$\%~CL. The blue line stands for the
potential reach on the sensitivity of SHiP experiment~\cite{SHiP:2020noy} assuming
five years of running.
The DUNE projection for
the sensitivity are shown in the red  and cyan lines
for DUNE on-axis and DUNE PRISM configurations respectively.

Concerning the DUNE limits, the projected sensitivity for the on-axis and the
DUNE PRISM configurations are both obtained using
{\tt MadDump} for signal simulation and  {\tt NuWro} for background simulation as
described in Sec. \ref{sec:model}.
In the analysis, it is assumed that the experiment will take data during seven years.
$i)$. For the DUNE on-axis analysis, 3.5 year in $\nu$-mode
and 3.5 years  in  $\overline{\nu}$-mode and $ii)$ for the DUNE PRISM analysis,
we combine the results of
on-axis position and different off-axis positions assuming that the detector takes data during
0.5 years in each mode ( $\nu$ and $\overline{\nu}$) at different axis configurations, corresponding
to vertical displacement of the detector from the beam
axis of $0$ m, $6$ m , $12$ m, $18$ m, $24$ m, $30$ m and $36$ m,
respectively. We also used {\tt MadDump} for signal simulations of SHiP and NO$\nu$A.

Additionally, in Fig.~\ref{plot:sensitivity} the dashed black and dashed green lines correspond
to the thermal relic density equal to the experimental value
 observed by Planck satellite, for one dark matter candidate (ODM) and two dark matter  candidates (TDM), respectively.
Finally, the cyan and dark blue stars represent the benchmark points displayed in Fig.~\ref{fig:ElectronEnergyMainText}.
  All in all, we conclude that by combining the on-axis
  with the off-axis data, the DUNE sensitivity
  increases around one order of magnitude
  in $\varepsilon^2$ for the explored mass range. Something that is
  interesting because it will,
eventually, allow us to probe an unexplored region
in the parameter space where the total
relic density is in accordance with
the observed one for ODM as well as for TDM candidates.
Finally, the cyan and dark blue stars represent
the benchmark points displayed in Fig.~\ref{fig:ElectronEnergyMainText}.
Notice that these two benchmark points,
so far, can only be probed by future DUNE PRIMS and SHiP experiments.
We emphasize that for the TDM case, bounds coming from the thermal
  relic density  and sensitivity allow us to explore
  a different region of the parameter space in comparison the ODM case.
   This result is novel and has not been reported in the literature.
It is worth to mention that our sensitivity analysis
results match those given in Refs.~\cite{DeRomeri:2019kic,Celentano:2020vtu}.
For the sensitivity analysis we follow the statistical description given in appendix~\ref{sec:statistical_analysis}.

\section{Conclusions}
\label{sec:conclusions}

In this work, we have presented a DM model that includes two stable scalar singlets and a new broken $U(1)_D$ symmetry. The model includes a massive dark photon that kinetically mixes with the SM photon. Thus, a DM candidate may communicate with SM particles through the photon-dark photon and with the other candidate through the dark photon portal. 
These two channels are very important in setting the relic abundance, but we find that DM conversion has a more profound effect because it dominates as long as it is viable. The result of this is that the heaviest DM particle converts into the lighter one, thus, only one DM particle remains. Nevertheless, the relic abundance phenomenology differs greatly from that of one DM particle. In fact, at the chosen benchmark point, for ODM the correct abundance has a diagonal shape in the $\epsilon^{2}$-$m_{\gamma^{\prime}}$ plane, whereas the TDM has a triangular shape. This shape allows for different values of $\epsilon^2$ at $m_{\gamma^{\prime}}\sim$ 0.06 GeV, and shifts the allowed $\epsilon^2$ for larger $m_{\gamma^{\prime}}$. We also find that currently this model can not be probed through DD experiments such as CRESST-III.

Besides, we systematically study the relevant background for DM signals produced by neutrino interactions ($\nu-e$) and $\nu_{e}$-CCQE at DUNE ND. We showed as less than 0.5$\%$ of $\nu_{e}$-CCQE events contribute to the total background for $E_{e}\theta_{e}^{2}<2$ MeV rad$^{2}$. We verify that the PRISM concept is critical in reducing the background events when the ND is moved for different off-axis positions. The possibility of cutting the $\nu_{e}$-CCQE background distributions and the advantages from the off-axis movement are crucial in the sensitivity studies.

Regarding the sensitivity analysis, we have found
that projected sensitivity for DUNE on-axis for seven
years of data taking will not have the potential
to explore the model, mainly because
the sensitivity limit at $90$\%~CL
will correspond to a region
which is already excluded by NA64.
The most important result comes from 
DUNE PRISM configuration that will improve
the on-axis sensitivity  by roughly
one order of magnitude in
$\varepsilon^2$, thus, allowing this
future configuration setup to test
an unexplored  region
of the parameter space of the model.
It turns out, that SHiP experiment, will be
able to cover an even bigger fraction of the parameter
space, because there are more control on the background.
  These experiments will open up a zone
  in the parameter space where the thermal
  relic density is fully explained
  for ODM as well as for TDM candidates, giving rise
  to the possibility to test the multiparticle
  nature of DM. The latter possibility, however,
  will depend on the ability of the experiment's detector
  to distinguishing signatures produced by
  different DM species at the detector.
Additionally, as a complementary analysis,
we  recast the current limits provided
by No$\nu$A  experiment, however,
it does not put any additional restriction
on the parameter space of the model,
to the point that, the NA64
bound  is slightly  more stringent than No$\nu$A.

\section{Acknowledgments}
\label{sec:acknow}
We  are  thankful  to Liliana Sánchez, Óscar  Zapata, Sandra Naranjo, Pedro Machado, Kevin Kelly,
Yuber Perez, Dimitry Zhuridov, Luca Buonocore  and the Maddump team for enlightening discussions.  A.B, J.S and G.P  are  supported
through  Universidad  EIA project grant II12019009.  A.C is supported by the project grant 616-2019 from Minciencias-Universidad Sergio Arboleda (Postdoctoral National Programme). For the final stages of this work, A.C received a financial support of the ``Flavor in the era of the LHC'' project,
grant PGC2018-102016-A-I00 funded by the Spanish MINECO.

\section{Appendix}
\label{sec:appendix}

\subsection{Total Expected Events for the Background}
\label{app:events}
Distributions for $\nu_{e}$-CCQE and $\nu-e$ backgrounds presented in Sec.~\ref{sec:signalbackground} have considered the normalizations computed from exposure on DUNE ND, cross sections and neutrino fluxes. The number of events for the $\nu_{e}$-CCQE and $\nu-e$ backgrounds are estimated via \cite{Rott_2017}

\begin{align}
 N_{\rm{events}}=A_{eff}\bar{F}\times(\text{Target Particles}),
\end{align}

where $\bar{F}$ is the relevant integrated neutrino flux. The effective area $A_{eff}$ of the detector is

\begin{align}
 A_{eff}=\sigma \frac{M_{D}}{M_{Ar}}N_{\text{POT}},
\end{align}

with $\sigma$ the cross section per nucleon for $\nu_{e}$-CCQE or $\nu-e$ scattering, $M_D$ (units of atomic mass) the mass of the detector, $M_{\text{Ar}}$ the mass of Argon (39.95 atomic mass), and $N_{\text{POT}}$ is the number of protons on target per year.

To determine the overall events distribution, we analyze $10^5$ CCQE $\nu+^{40}\text{Ar}$ and $\nu-e$ events generated in \texttt{NuWro}. To use the $^{40}$Ar spectral function, we set \texttt{nucleus$\_$target=2} (Relativistic Fermi Gas)
and \texttt{sf$\_$method=1} (realistic density profile), and \texttt{cc$\_$axial$\_$mass=1030} MeV in the \texttt{NuWro} parameters-initialization file. When the target nucleon is bound in the parent nucleus, the Relativistic Fermi Gas Model (RFG) describes the initial nuclear state; the neutrons of Argon in the case of $\nu_{e}$-CCQE. Besides, the final state hadrons produced at the primary neutrino collision undergo non-perturbative effects of strong interactions inside the target nucleus. In this case, the absence of well-defined and complete models makes the treatment of these nuclear effects a relevant source of systematic uncertainty in the Montecarlo simulations~\cite{Golan:2012rfa,Alvarez-Ruso:2017oui}. All other commands in the parameter card are the default values.

By using official neutrino fluxes at the near detector facility for the on-axis and on off-axis positions and running in the – neutrino and antineutrino – modes \cite{LFFluxesDUNE} (design with a proton beam of 120 GeV) and the average cross-section given by \texttt{NuWro}, we compute the number of expected neutrino-electron scattering and $\nu_e$-CCQE events in the DUNE liquid argon near detector (listed in Tab \ref{tab:numode}).

\begin{table}[ht]
\begin{center}
\begin{tabular}{|c|c|c|c|c|c|c|c|c|}\hline
\hline
\multicolumn{8}{ |c| }{\bf DUNE ND (${\nu}$ and $\bar{\nu}$ modes)} \\
\hline \hline
\bf Channel&\bf On axis&\bf 6 m&\bf 12 m   & \bf 18 m & \bf 24 m  &\bf 30 m & \bf 36 m \\\hline \hline
\multirow{2}{*}{$\nu_{\mu}e\to{\nu}_{\mu}e$} &115  & 78 & 27 & 13 & 7 & 4   & 3 \\  
& 12 & 10 & 5 & 3 & 2 & 1 & 1  \\ \hline
\multirow{2}{*}{$\nu_{e}e\to {\nu}_{e}e$} &  9 & 6 & 4 & 2 & 1 & 1   & 1 \\\ 
& 3 & 2 & 2 & 1 & 0 & 0 & 0 \\ \hline
\multirow{2}{*}{$\bar{\nu}_{\mu}e\to \bar{\nu}_{\mu}e$} &  8 & 6 & 3 & 2 & 1 & 1   & 1 
\\
& 82 & 45  & 20  & 9 & 5 & 3 & 2 \\ \hline
\multirow{2}{*}{$\bar{\nu}_{e}e\to\bar{\nu}_{e}e$} &  1 & 1 & 1 & 0 & 0 & 0   & 0 \\ 
& 3 & 2  & 1  & 1 & 0 & 0 & 0 \\
\hline
\multirow{2}{*}{\bf{Total} $\nu e\to{\nu}e$} & 133 & 91 & 35 & 17 & 9 & 6   & 5    \\ 
& 100 & 59  & 28  & 14 & 7  & 4 & 3 \\
\hline \hline
\multirow{2}{*}{\bf{Total} $\nu_e$-CCQE} &  3003 & 2355 & 1572 & 1067 & 750 & 550   & 412 \\ 
& 676 & 595  & 476  & 357 & 272 & 219 & 178 \\
\hline 
\multirow{2}{*}{$\nu_e$-CCQE$^{*}$}  & 14 & 10 & 6 & 3 & 2 & 1 & 0 \\ 
& 3 & 3  & 2  & 1 & 1 & 0 & 0 \\
\hline 
\end{tabular}
\end{center}
\caption{ Expected number of $\nu-e$ and $\nu_{e}$-CCQE events (per ton per year) in DUNE ND for on-axis and six off-axis positions. Events numbers in $\nu$ ($\bar{\nu}$) mode are in top (bottom) row. Fluxes consider $0.125<E_{\nu}<10.125$ GeV. The last row for $\nu_{e}$-CCQE$^{*}$ has the events accepted  by the kinematical cut $E_{e}\theta^2_{e}<2$ MeV rad$^{2}$. The accepted events represent less than 0.5$\%$ of the total $\nu_{e}$-CCQE background.}
\label{tab:numode}
\end{table}

 \subsection{Statistical Analysis}
   \label{sec:statistical_analysis}
 In this section, we provide a description of the two different approaches used for 
 building the statistical test applied in the analysis, ultimately  both of them leading to equivalent results.

\begin{itemize}

\item {\bf Log-likelihood}: Closely following  the
  statistical test proposed in Refs~\cite{DeRomeri:2019kic, Breitbach:2021gvv},
  we build a Poisson log-likelihood function to compare the signal plus background
  hypothesis against the background-only hypothesis\footnote{
    For DUNE, events for signal and background are both simulated.
    In the case of NO$\nu$A and SHiP, only signal events are simulated, the background
    are taking from the collaboration report~\cite{SHiP:2020noy,Bian:2017axs,Acero:2019qcr,Acero:2019ksn}.}.
  For the signal as well as the background events, we consider
  the electron recoil energy distribution, with 
  a total of $40$ bins  for  energies in the range
  $[0-10]$~GeV.
  The modified Poissonian likelihood function per energy bin $j$ and detector position $i$ reads
  \begin{eqnarray}{\label{eq:poisson_likelihood}}
    \mathcal{L}_{ij}(\mu) = \dfrac{(w_{ij} )^{d_{ij}}e^{-(w_{ij})} }{d_{ij}!} \quad
    \text {with}  \quad  w_{ij} = A f_i  ( \mu N_{ij}^{\phi} + N_{ij}^{\nu} )~,
  \end{eqnarray}
  where $N_{ij}^{\nu}$,  $N_{ij}^{\phi}$ and  $d_{ij}$ stands for number of background events, signal events, and observed
  events respectively,  for each detector position $i$ and
  energy bin $j$. The terms $A$ and $f_i$ are nuisance parameters for overall systematic uncertainty and
  detector positions respectively. Finally, $\mu$ is a parameter which is set to 1 for signal plus background hypothesis
  and set to $0$ for background-only hypothesis. Since for DUNE ND and SHiP
  there is not observed data, we assume $d_{ij}= N_{ij}^{\nu}$  in
  the analysis.
  The binned test statistics is defined as
  \begin{eqnarray}{\label{eq:test_statistic_1}}
  q = -2\Delta L =  \sum_{i=1}^{n_{\rm pos}} \Bigg [ \sum_{j=1}^{n_{\rm bins}} -2 \log \Bigg ( \dfrac{\mathcal{L}_{ij}(\mu=1)}{\mathcal{L}_{ij}(\mu=0)} \Bigg )
    + \dfrac{(f_i -1)^2 }{\sigma_{f_i}^2} \Bigg ]  + \dfrac{(A -1)^2 }{\sigma_{A}^2}~,
  \end{eqnarray}
  where we assume $\sigma_{f_i}=1\%$ and $\sigma_{A}=10\%$, 
  ${n_{\rm bins}}$  being
  the total number of bins and  ${n_{\rm pos}}$ the total number of detector positions.
  A minimization of $q$ respect to the nuisance parameters is performed in order to find them.
    By proceeding in such a way a total of  $n_{\rm pos} + 1$ non-linear equations arises, allowing to obtain
    the nuisance parameters~\cite{Conway:2011in}.  
  The sensitivity plots at  $90$\% CL demands  $q>4.61$ for two
  degrees of freedom.

\item{\bf Unbinned analysis}: The second possibility is to
  define the significance  as \cite{SHiP:2020noy,Zyla:2020zbs}:
  \begin{eqnarray}{\label{eq:test_statistic_2}}
  Z = \dfrac{N^{\phi}}{\sqrt{N^{\nu} + \displaystyle \sum_{i \in (\nu_e, \nu_{\mu},\overline{\nu}_e,\overline{\nu}_{\mu} )} (k_iN^{\nu}_{i})^2 } }~,
  \end{eqnarray}
  where $N^{\phi}$ corresponds to the total number of signal events, $N^{\nu}$ the total background events,
  $N^{\nu}_{i}$ the background events per each $\nu$-flavor background, and  the  $k_i$ factors stand for the
  overall systematic uncertainty for each $\nu$-flavor background.
  For the sensitivity analysis,  $k_i = 10\%$ for DUNE and $k_i = 5\%$ for SHiP are used, while for
  NO$\nu$A, no uncertainties are considered. 
  The sensitivity at $90$\% CL is obtained by setting  $Z\geq 1.64$.
\end{itemize}

\newpage
\bibliographystyle{apsrev4-1long}
\bibliography{biblioU1}







\end{document}